\documentclass[journal]{IEEEtran}
\usepackage{amsmath,amsfonts}
\usepackage{algorithmic}
\usepackage{array}
\usepackage[caption=false,font=normalsize,labelfont=sf,textfont=sf]{subfig}
\usepackage{textcomp}
\usepackage{stfloats}
\usepackage{url}
\usepackage{verbatim}
\usepackage{graphicx}
\usepackage{bm}
\usepackage{multirow}
\usepackage{makecell}
\usepackage{booktabs}
\usepackage{color}
\usepackage{adjustbox}
\usepackage{cite}
\usepackage{xcolor}
\usepackage{colortbl}
\usepackage{algorithm,algorithmic}
\usepackage{threeparttable}
\usepackage{booktabs,makecell,subcaption}
\newcommand{\best}[1]{\textbf{#1}}
\newcommand{\second}[1]{\underline{#1}}
\usepackage{balance}

\newcommand{\RevisedText}[1]{
\textcolor{black}{#1}
}

\begin{document}
\bstctlcite{settingbib}

\title{WiFo-MiSAC: A Wireless Foundation Model for Multimodal Sensing and Communication Integration via Synesthesia of Machines (SoM)}
 
\author{\IEEEauthorblockN{
		Xuanyu Liu,~\IEEEmembership{Graduate Student Member,~IEEE,}~Shijian Gao,~\IEEEmembership{Member,~IEEE,}~Boxun Liu,~\IEEEmembership{Graduate Student Member,~IEEE,}~Xiang Cheng,~\IEEEmembership{Fellow,~IEEE,}~and Liuqing Yang,~\IEEEmembership{Fellow,~IEEE}
        }
\thanks{
Xuanyu Liu, Boxun Liu, and Xiang Cheng are with the State Key Laboratory of Photonics and Communications, School of Electronics, Peking University, Beijing 100871, China (e-mail: xyliu25@stu.pku.edu.cn; boxunliu@stu.pku.edu.cn; xiangcheng@pku.edu.cn).
			
Shijian Gao is with the Internet of Things Thrust, The Hong Kong University of Science and Technology (Guangzhou), Guangzhou 511400, China (e-mail: shijiangao@hkust-gz.edu.cn).
		
Liuqing Yang is with the Internet of Things Thrust and Intelligent Transportation Thrust, The Hong Kong University of Science and Technology (Guangzhou), Guangzhou 511400, China (e-mail: lqyang@ust.hk).
}} 

\maketitle

\begin{abstract}
Current learning-based wireless methods struggle with generalization due to the fragmented processing of communication and sensing data. WiFo-MiSAC addresses this as a task-agnostic foundation model that tokenizes heterogeneous signals into a unified space for self-supervised pre-training. A shared-specific disentangled mixture-of-experts (SS-DMoE) architecture is employed to explicitly model modality-shared and modality-specific representations, facilitating interaction without cross-modal interference. By combining masked reconstruction with contrastive alignment, the model achieves state-of-the-art performance across downstream tasks, including beam prediction and channel estimation. Experimental results demonstrate robust few-shot adaptation and seamless integration of new modalities, positioning WiFo-MiSAC as a scalable backbone for future integrated sensing and communication systems.
\end{abstract}



\section{Introduction}

\begin{figure*}[htbp]
    \centering
    \vspace{-1em}
    \includegraphics[width=\linewidth]{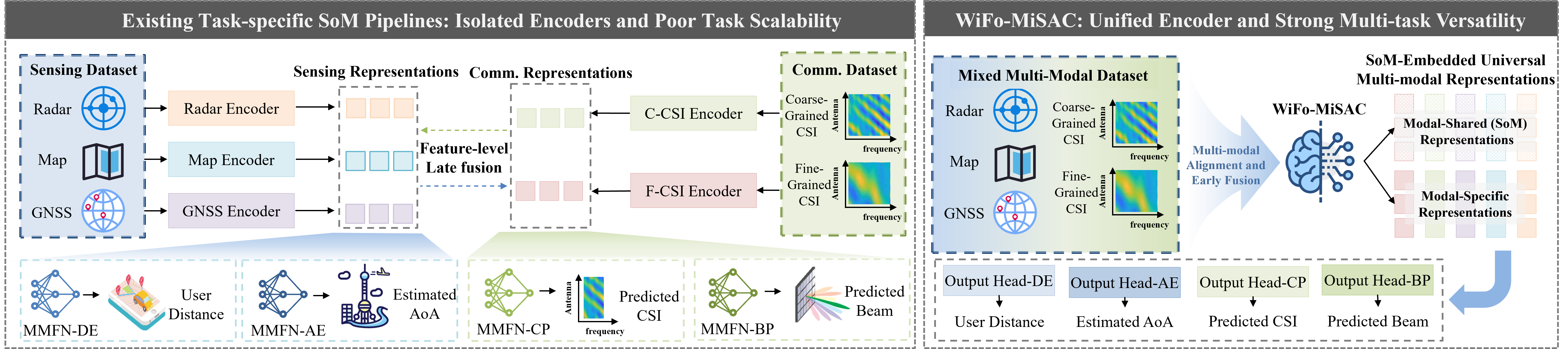}
\caption{\RevisedText{Comparison between conventional task-specific SoM pipelines and the proposed WiFo-MiSAC framework.}}
    \label{fig:workflow}
    \vspace{-1em}
    \end{figure*} 

\textcolor{black}{Researchers widely view} the convergence of communication, sensing, computing, and intelligence \textcolor{black}{as a defining trend} of sixth-generation (6G) wireless systems.     
Within this vision, the intelligent fusion of communication and multimodal sensing is expected to enable emerging applications
such as Vehicle-to-everything (V2X) \cite{cheng2022integrated, liu2022integrated}, low-altitude aerial networks \cite{jiang2025integrated}, and embodied intelligent networked agents.
To systematically characterize the design paradigm and underlying mechanisms of communication--multimodal sensing fusion, Cheng \emph{et al.} introduced the notion of \emph{Synesthesia of Machines} (SoM) \cite{cheng2023intelligent}. Guided by this perspective, recent studies have investigated how multimodal sensing signals, such as red--green--blue (RGB) images, light detection and ranging (LiDAR), and radar, can reduce training and pilot overhead while improving task performance in beam alignment~\cite{zhang2024integrated}, channel prediction~\cite{yang2020deep}, channel estimation~\cite{mundlamuri2023sensing}, and blockage identification~\cite{demirhan2022radar2}, with effectiveness validated on real-world datasets.

Despite this progress, the prevailing \emph{SoM system design} paradigm remains largely \emph{task- and scenario-specific}. In a typical pipeline, modality encoders and fusion operators are manually designed for a specific system configuration and downstream objective, and the resulting model is trained via supervised learning. 
Representative studies leverage auxiliary sensing modalities, such as LiDAR\cite{klautau2019lidar}, cameras \cite{alrabeiah2020millimeter}, and positional information \cite{charan2022towards}, to aid mmWave beam selection and blockage awareness in specific settings.
While effective within the target scenario, such designs exhibit three structural limitations:
\begin{itemize}
    \item \textbf{Architectures are tightly coupled to specific signal dimensions:} new sensor configurations require costly re-engineering. This lack of scalability necessitates frequent retraining.
    \item \textbf{Late fusion restricts fine-grained cross-modal alignment:} models struggle to capture complex sensing-communication correlations. This results in negative transfer during signal misalignment.
    \item \textbf{Narrow training distributions lead to brittle performance:} models fail under domain shifts or hardware noise. Such sensitivity prevents robust generalization in dynamic environments.
\end{itemize}

Recently, the \emph{foundation model} (FM) paradigm \cite{wang2025sam, zhang2025dmae, cheng2025foundation, cheng2026large, wang2026consisformer, wang2025filter, yang2025wirelessgpt} has offered a unifying perspective for physical-layer intelligence: large-scale self-supervised pre-training learns general-purpose representations that can be efficiently adapted to diverse downstream tasks with lightweight fine-tuning. As illustrated in Table~\ref{tab:model_comparison}, representative examples include LWM \cite{alikhani2024large}, which employs Transformer-based self-supervision to produce transferable channel embeddings for downstream tasks such as scenario classification and beamforming, and WiFo \cite{liu2024wifo}, which pioneers a pre-training scheme tailored to channel prediction and exhibits strong zero-shot generalization. Collectively, these results indicate that scaling pre-training on diverse CSI corpora yields robust, configuration-transferable representations, alleviating the scalability bottlenecks of task-specific models.

However, most existing wireless FMs remain fundamentally \emph{unimodal}, typically operating on CSI \cite{liu2025wifo, liu2025foundation} or its transformations \cite{mashaal2026iqfm}, and therefore do not natively fit the prevalent SoM input form that couples communications with multimodal sensing. While one may incorporate vision, radar, or map priors via post-hoc alignment or lightweight adapters, such retrofitting usually introduces additional cross-modal plumbing and handcrafted fusion choices, offers limited portability across heterogeneous system configurations and downstream objectives, and degrades noticeably when modalities are added, missing, or perturbed. More importantly, adapter-based coupling often constrains the model to shallow, late-stage interactions, making it difficult to capture the deep and fine-grained dependencies between communication and sensing modalities that are essential for complex SoM tasks, ultimately limiting performance. These limitations motivate \emph{native} multimodal wireless foundation models that can provide unified, flexible, and efficient support for a broad spectrum of wireless communication--sensing tasks.

Several concurrent works adopt \emph{contrastive learning} with \emph{dual-tower encoders} to align sensing and communication embeddings. For example, WiFo-M$^2$ \cite{zhang2026wifo} employs contrastive-style pre-training to map environmental sensing to ``channel-relevant'' features, serving as a plug-and-play enhancement module for various physical-layer actions. WMFM \cite{farzanullah2025wireless} similarly aligns visual and channel representations via contrastive objectives and demonstrates transfer gains in tasks such as localization and LoS/NLoS classification. 
\RevisedText{Nevertheless, this line of solutions has two notable limitations. First, when contrastive learning is used as the primary pre-training signal, its effectiveness heavily depends on high-quality paired cross-modal data and carefully constructed inter-sample relations, which may limit its robustness under imperfect sensing--communication correspondence. Second, dual-tower alignment primarily enforces \emph{global embedding proximity}, which is insufficient for capturing fine-grained communication--sensing correlations at the token and spatiotemporal levels \cite{yan2025frcl}; moreover, performance can degrade sharply under missing modalities or severe cross-modal misalignment, sometimes failing structurally.}

Departing from prior designs, we develop a unified multimodal Transformer backbone that flexibly supports variable modality sets for both input and output, as shown in Fig.\ref{fig:workflow}. Such a unified architecture reduces the reliance on strictly aligned cross-modal pairs, enabling pre-training on richer weakly aligned or partially unpaired data at a larger scale, and naturally improves robustness to modality expansion, dropouts, and missing inputs at inference time. Inspired by SoM, we explicitly factorize multimodal features into two complementary components: (i) \emph{modality-specific} representations that preserve modality-specific structures, and (ii) a \emph{modality-shared} representation, termed the \emph{SoM representation}, that captures unified environmental semantics jointly reflected across modalities. Building on this decomposition, we propose the Shared--Specific Disentangled Mixture-of-Experts (SS-DMoE) architecture in which \emph{modality-specialized expert groups} collaborate with a \emph{SoM expert group}, enabling explicit and interpretable decoupling of specific versus shared factors; this mitigates modality conflicts often observed in late-stage feature fusion and strengthens the model's ability to capture deep cross-modal interactions required by complex SoM tasks. Furthermore, we devise a self-supervised pre-training scheme that combines masked reconstruction with contrastive objectives, which promotes fine-grained cross-modal alignment beyond global embedding matching, thereby improving fusion quality and yielding more effective SoM representations for downstream adaptation. The core contributions of this paper are summarized below:

\begin{itemize}
    \item \RevisedText{We develop WiFo-MiSAC, a native multimodal wireless foundation model for integrated sensing and communication, equipped with a customized SS-DMoE architecture. By representing CSI, FMCW radar, and environmental maps as heterogeneous physical observations of the same propagation environment, WiFo-MiSAC explicitly models modality-shared environmental factors from modality-specific measurement structures, enabling environment-adaptive cross-modal sharing while preserving modality-intrinsic characteristics.}

    \item \RevisedText{We construct a large-scale multimodal sensing--communication dataset and propose a hybrid self-supervised pre-training strategy tailored to practical wireless data availability. The dataset contains more than 200k CSI--radar--map triplets covering diverse propagation scenarios and system configurations. The pre-training objective combines cross-modal masked reconstruction and CSI-anchored contrastive alignment, allowing WiFo-MiSAC to exploit both complete paired samples and incomplete or partially unpaired multimodal observations without relying exclusively on strictly synchronized data.}

    \item \RevisedText{We demonstrate that WiFo-MiSAC serves as a general-purpose multimodal backbone for wireless sensing and communication. Comprehensive experiments show that a single pre-trained model supports zero-shot channel reconstruction, unified adaptation to multiple communication and sensing downstream tasks, robust inference under missing modalities, and efficient extension to new modalities. These results confirm the effectiveness of the proposed wireless-specific representation architecture and its potential as a scalable backbone for practical integrated sensing and communication systems.}
\end{itemize}

\begin{table*}[t]
\centering
\vspace{-1em}
\setlength{\tabcolsep}{5pt}
\renewcommand{\arraystretch}{1.15}
\caption{\textcolor{black}{Comparative analysis of wireless foundation models in terms of training strategies and model capabilities.}}

\resizebox{\textwidth}{!}{%
\begin{tabular}{|c|c|c|c|c|c|c|c|c|c|}
\hline
\multirow{2}{*}{Method} 
& \multirow{2}{*}{Modalities} 
& \multirow{2}{*}{\makecell{Fusion Strategy}} 
& \multirow{2}{*}{\makecell{Training\\Objectives}} 
& \multicolumn{2}{c|}{\makecell{Input Support}} 
& \multicolumn{2}{c|}{\makecell{Scalability \& Robustness}} 
& \multicolumn{2}{c|}{Generalization} \\
\cline{5-10}
& & & 
& \makecell{Unimodal} 
& \makecell{Multimodal} 
& \makecell{New Modality} 
& \makecell{Missing Modality} 
& \makecell{Cross-\\Scenario} 
& \makecell{Cross-\\Configuration} \\
\hline

LWM\cite{alikhani2024large}
& CSI 
& / 
& MM 
& \checkmark 
& $\times$ 
& $\times$ 
& $\times$ 
& \checkmark 
& $\times$ \\
\hline

WiFo\cite{liu2024wifo}
& CSI 
& / 
& MM 
& \checkmark 
& $\times$ 
& $\times$ 
& $\times$ 
& \checkmark 
& \checkmark \\
\hline

WiFo-M$^2$\cite{zhang2026wifo}
& CSI+RGB+LiDAR 
& \makecell{Late fusion} 
& CL 
& $\times$ 
& \checkmark 
& $\times$ 
& $\times$ 
& \checkmark 
& \checkmark \\
\hline

WMFM\cite{farzanullah2025wireless}
& CSI+RGB 
& \makecell{Late fusion} 
& CL 
& $\times$ 
& \checkmark 
& $\times$ 
& $\times$ 
& \checkmark 
& $\times$ \\
\hline

PHYFM\cite{yazdnian2026multi}
& CSI+Map 
& \makecell{Early fusion} 
& MM 
& $\times$ 
& \checkmark 
& $\times$ 
& $\times$ 
& \checkmark 
& $\times$ \\
\hline
\rowcolor{gray!15}
\textbf{WiFo-MiSAC} 
& \textbf{CSI+Radar+Map}
& \textbf{Early fusion} 
& \textbf{CL+MM} 
& \textbf{\checkmark} 
& \textbf{\checkmark} 
& \textbf{\checkmark} 
& \textbf{\checkmark} 
& \textbf{\checkmark} 
& \textbf{\checkmark} \\
\hline

\end{tabular}%
}
\begin{tablenotes}
\footnotesize
\item Note: CL denotes contrastive learning, and MM denotes masked modeling.
\end{tablenotes}
\label{tab:model_comparison}
\end{table*}

\section{\RevisedText{Related Work}}
\subsection{\RevisedText{Multimodal Integration of Sensing and Communication}}
\RevisedText{Multimodal sensing--communication integration incorporates environmental observations, such as radar, cameras, LiDAR, maps, and positioning information, into wireless signal processing. These complementary modalities provide geometric, mobility, and blockage-related information that is difficult to obtain from communication signals alone, and have been applied to beam prediction, channel prediction and estimation, blockage identification, localization, and environmental perception \cite{zhang2024integrated,yang2020deep,mundlamuri2023sensing,demirhan2022radar2}}.

\RevisedText{Most existing methods are developed for a predefined task and modality configuration. A common design is to encode each modality independently and combine the resulting features through concatenation, attention, or a task-specific fusion module. Although effective in their target settings, these models are often closely tied to particular signal dimensions, sensing configurations, and downstream objectives. Changes in the modality set or system configuration may therefore require additional model adaptation or retraining.}

\RevisedText{Another practical challenge arises from the heterogeneous acquisition processes of sensing and communication signals. Different modalities may operate at different sampling rates, coverage ranges, and synchronization accuracies, making fully synchronized multimodal observations difficult to obtain at scale. This motivates a unified representation-learning framework that can accommodate variable modality combinations, exploit both complete and incomplete observations, and transfer across communication and sensing tasks.}

\subsection{\RevisedText{Mixture-of-Experts for Multimodal Fusion}}
\RevisedText{Mixture-of-experts architectures have been widely adopted in multimodal learning to improve representation capacity while encouraging expert specialization. VLMo introduces modality-specific experts within a unified vision--language Transformer, while EVE combines modality-aware sparse experts with masked multimodal pre-training \cite{bao2022vlmo, chen2024eve}. AV-MoE and BR-MoE further explore sparse expert activation and dynamic routing for heterogeneous multimodal inputs \cite{cheng2024mixtures,meng2025br}. Other fusion methods, such as MSTN and SelectFusion, model cross-modal interaction through sparse attention or reliability-aware feature selection \cite{chen2022learning,song2022multimodal}.}

\RevisedText{These approaches mainly focus on semantic modalities or task-oriented sensor fusion. In multimodal wireless systems, CSI, FMCW radar, and environmental maps provide different measurements of the same propagation environment. Their observations are correlated through scene geometry, blockage, scattering, and multipath propagation, while each modality retains its own signal structure, resolution, and noise characteristics. Moreover, the correspondence among these modalities varies with the propagation environment and system configuration.}

\RevisedText{This physical coupling requires a representation model that preserves modality-dependent information while extracting transferable environmental factors. WiFo-MiSAC addresses this requirement through SS-DMoE, where shared SoM experts capture propagation-related commonality and modality-specific experts retain signal-dependent characteristics. Environmental context derived from map tokens is further incorporated into the shared-expert routing, allowing cross-modal interaction to adapt to scene-dependent propagation conditions.}

\section{System Model}
This section introduces a unified sensing-enabled cellular framework where a multi-antenna base station (BS) is co-located with auxiliary sensors like FMCW radar. To support diverse 6G functionalities, from channel estimation to AoA prediction, a multimodal foundation model is deployed to process heterogeneous inputs for a suite of downstream tasks. Unlike traditional sensing-aided communication that treats radar as mere side information, this approach enables bidirectional fusion. By jointly modeling sensing and communication streams, the system captures mutual signal benefits and addresses critical sensing-layer objectives such as localization and mapping. This formulation establishes a scalable, unified backbone for intelligent 6G interfaces.

\subsection{Modalities}
Three modalities are considered in this work: CSI, FMCW radar, and map. CSI is the cornerstone of the physical layer and the primary object for communication optimization, as it directly characterizes the propagation channel for tasks such as estimation, prediction, and beamforming. Radar is particularly attractive for practical deployment due to its robustness under challenging illumination and weather conditions, as well as its compact and cost-effective form factor compared with high-end optical sensors. The map serves as a complementary, heterogeneous sensing source that provides a global structural view of the environment: it can mitigate radar degradation in NLoS conditions caused by occlusions, offer an additional observation perspective beyond instantaneous measurements, and improve cross-scenario generalization by capturing relatively stable scene geometry. Notably, while our experiments focus on these three modalities, our proposed \emph{architecture} is not restricted to them: thanks to a unified modality tokenization and a shared backbone, additional sensing streams (e.g., RGB images, depth, or LiDAR point clouds) can be seamlessly integrated with minimal modification.

\subsection{Communication model}
We consider a single-user multiple-input multiple-output orthogonal frequency division multiplexing (MIMO--OFDM) link where the BS and UE are equipped with $N_{\mathrm{t}}$ transmit and $N_{\mathrm{r}}$ receive antennas, respectively, over $N_{\mathrm{sc}}$ subcarriers. The downlink channel is modeled by a clustered multipath representation. At carrier frequency $f_0$, the frequency-domain channel vector can be written as
\begin{equation}\label{CSI}
\bm{h}(f)=\sum_{m=1}^{M}\sum_{p=1}^{P_m}\beta_{m,p}\,e^{-j2\pi f \tau_{m,p}}\,e^{j\Phi_{m,p}}\,
\bm{a}(\theta_{m,p},\phi_{m,p}),
\end{equation}
where $M$ is the number of clusters and $P_m$ is the number of rays in the $m$-th cluster. Here, $\beta_{m,p}$, $\tau_{m,p}$, and $\Phi_{m,p}$ denote the complex gain, delay, and phase of the $(m,p)$-th path, respectively, and $\bm{a}(\theta_{m,p},\phi_{m,p})$ is the array steering vector with azimuth $\theta_{m,p}$ and elevation $\phi_{m,p}$.


\subsection{Radar model}
The BS is equipped with an FMCW radar. At sensing interval $t$, the radar returns a complex baseband data cube
 $\mathcal{R}_t \in \mathbb{C}^{N_{\mathrm{rx}} \times N_{\mathrm{chirp}} \times N_{\mathrm{samp}}}$, indexed by receive antenna, chirp, and ADC sample. Instead of extracting sparse point clouds or directly tokenizing the full 3D cube, we convert $\mathcal{R}_t$ into two dense 2D representations. The key idea is to apply a 2D FFT along the two dimensions relevant to the target attributes, and then average over the remaining dimension to obtain a stable intensity map.

\textbf{Range--angle (R--A) map.}
We apply a 2D FFT over the receive antenna dimension and the ADC sample dimension, and then average the magnitude over chirps:
\begin{equation}
\mathbf{R}^{\mathrm{RA}}_t
=
\frac{1}{N_{\mathrm{chirp}}}
\sum_{c=1}^{N_{\mathrm{chirp}}}
\left|
\mathcal{F}_{2\mathrm{D}}
\!\left(\mathcal{R}_t[:,c,:]\right)
\right|
\in \mathbb{R}^{N_{\mathrm{rx}}\times N_{\mathrm{samp}}}.
\end{equation}
\textcolor{black}{Here, $\mathcal{F}_{2\mathrm{D}}$ denotes the 2D FFT along the receive-antenna and ADC-sample dimensions, yielding angle and range bins. Following \cite{demirhan2022radar}, zero-padding is used to improve angular resolution. Besides increasing angular granularity, this process standardizes inputs across different devices, facilitating model adaptation to varying antenna configurations. These refined angle bins provide richer radar tokens for capturing detailed features with only a marginal increase in computational overhead.}

\textbf{Range--velocity (R--V) map.}
Similarly, we apply a 2D FFT over the chirp dimension and the ADC sample dimension, and then average the magnitude over receive antennas:
\begin{equation}
\mathbf{R}^{\mathrm{RV}}_t
=
\frac{1}{N_{\mathrm{rx}}}
\sum_{m=1}^{N_{\mathrm{rx}}}
\left|
\mathcal{F}_{2\mathrm{D}} 
\!\left(\mathcal{R}_t[m,:,:]\right)
\right|
\in \mathbb{R}^{N_{\mathrm{chirp}}\times N_{\mathrm{samp}}}.
\end{equation}
Here, $\mathcal{F}_{2\mathrm{D}}$ denotes the 2D FFT along chirps and ADC samples, yielding Doppler bins and range bins.

The resulting $\mathbf{R}^{\mathrm{RA}}_t$ and $\mathbf{R}^{\mathrm{RV}}_t$ preserve informative radar structure while remaining token-efficient, and are used as radar inputs to the unified multimodal backbone.

\subsection{Map model}
\label{sec:map_model}

We assume the BS has access to a local map within a radius of $100$\,m, which we represent by two rasterized tensors: a bird's-eye-view (BEV) map and a height map. The RGB BEV map is given by
\begin{equation}
\mathbf{M}^{\mathrm{BEV}} \in \mathbb{R}^{ H_{\mathrm{m}}\times W_{\mathrm{m}}\times3},
\end{equation}
which provides a top-down view of the scene layout (e.g., road geometry, building footprints, and other structural cues visible in the map rendering) and thus implicitly encodes the planar distribution of dominant scatterers. To complement the planar structure with vertical geometry, we additionally construct a height map
\begin{equation}
\mathbf{M}^{\mathrm{H}} \in \mathbb{R}^{H_{\mathrm{m}}\times W_{\mathrm{m}}\times1},
\end{equation}
where each grid cell stores its elevation value on the same spatial lattice. Together, $\mathbf{M}^{\mathrm{BEV}}$ and $\mathbf{M}^{\mathrm{H}}$ provide a compact yet informative description of 3D scene geometry, offering a richer proxy of the spatial distribution of potential scattering structures than either view alone.



\section{Method}
\label{sec:method}

\subsection{Motivation and Overview}
Wireless sensing and communication are increasingly reliant on multimodal observations (e.g., CSI and radar). A prevalent paradigm is \emph{late fusion}, where each modality is encoded independently and fused only at the final stage. While simple and effective in practice, late fusion often suffers from three limitations: (i) it underutilizes fine-grained cross-modal interactions; (ii) it lacks an explicit mechanism to model inter-modality relevance; and (iii) it degrades markedly under scene shifts, since the absence of explicit environmental embeddings prevents the model from extracting dynamically adaptive features in response to scene-dependent multipath effects.
To address these issues, we propose a unified wireless foundation model for multimodal sensing and communication
integration, termed \textbf{WiFo-MiSAC}. Specifically, we prepend map tokens as a context prefix and represent all modalities in a unified token space, enabling early and fine-grained fusion, thereby enhancing scene-aware multimodal integration. Inspired by SoM, we further develop a \textbf{SS-DMoE} that explicitly disentangles modality-shared representations from modality-specific representations. 
Moreover, we introduce an \emph{environment-context-aware router} that selects experts conditioned on a compact scene representation pooled from the map tokens, enabling \emph{scene-adaptive inference}. The overall network components are illustrated in Fig.~\ref{fig:network}. In the following sections, we detail the architecture of WiFo-MiSAC and its corresponding training pipeline.


\begin{figure}[htbp]
    \centering
    \vspace{-1em}
    \includegraphics[width=1\linewidth]{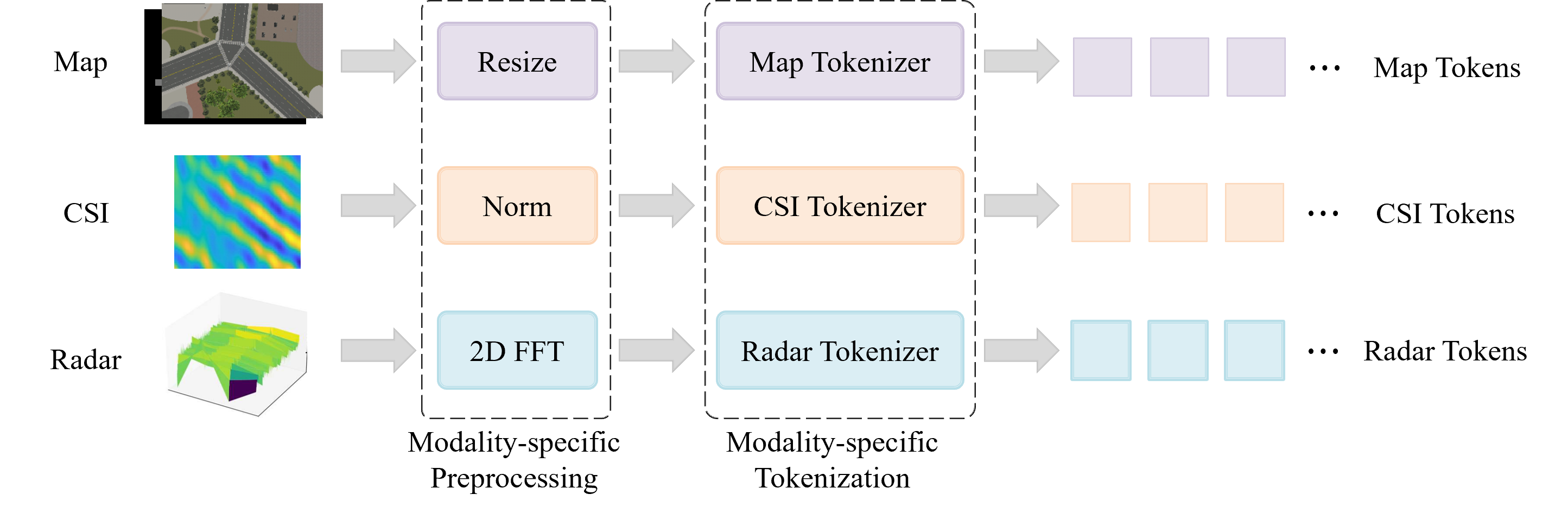}
    \caption{\textcolor{black}{An illustration of data pre-processing and modal-specific tokenizer.}}
    \label{fig:tokenizer}
    \vspace{-1em}
    \end{figure} 

\subsection{Data Pre-processing and Modal-specific Tokenizer}
\label{sec:tokenizer}

Let $\mathcal{M}=\{\texttt{csi},\texttt{map},\texttt{radar}\}$ denote the modality set. Each sample provides synchronized observations $\mathbf{X}^{(m)}$ for $m\in\mathcal{M}$. For CSI, the channel measured at the BS is $\mathbf{X}_0^{(\texttt{csi})}\in\mathbb{C}^{N_a\times N_{sc}}$, where $N_a=N_tN_r$ is the antenna dimension and $N_{sc}$ is the number of subcarriers. We convert it to a real-valued tensor $\mathbf{X}^{(\texttt{csi})}\in\mathbb{R}^{N_a\times N_{sc}\times 2}$ by concatenating the real and imaginary parts along a new dimension.

For radar, the R--A and R--V maps are already real-valued representations. We therefore concatenate them directly along the channel dimension before tokenization:
\[
\mathbf{X}^{(\texttt{radar})}=\mathrm{cat}\!\left(\mathbf{R}^{\mathrm{RA}}_t,\mathbf{R}^{\mathrm{RV}}_t\right).
\]
For map, we similarly concatenate the BEV map and height map along the channel dimension:
\[
\mathbf{X}^{(\texttt{map})}=\mathrm{cat}\!\left(\mathbf{M}^{\mathrm{BEV}},\mathbf{M}^{\mathrm{H}}\right).
\]
This preserves complementary structure while keeping a unified 2D tokenizer.

To make heterogeneous modalities compatible with a Transformer backbone, we employ a unified non-overlapping 2D patch embedding scheme \cite{dosovitskiy2020image}, as illustrated in Fig.~\ref{fig:tokenizer}. For each modality $m$, we partition the input $\mathbf{X}^{(m)}\in\mathbb{R}^{H_m\times W_m\times C_m}$ into patches of size $P^m_h\times P^m_w$, resulting in
\begin{equation}
N_m=\frac{H_m}{P^m_h}\cdot \frac{W_m}{P^m_w}.
\label{eq:num_tokens_new}
\end{equation}
Each patch is flattened and mapped to a $d$-dimensional token via a modality-specific linear projection:
\begin{equation}
\mathbf{z}^{(m)}_i
=
\mathbf{W}^{(m)}\,\mathrm{vec}\!\left(\mathbf{X}^{(m)}_i\right)+\mathbf{b}^{(m)},
\quad i=1,\ldots,N_m,
\label{eq:patch_embed_new}
\end{equation}
where $\mathbf{W}^{(m)}$ and $\mathbf{b}^{(m)}$ are learnable parameters. The resulting token sequence is denoted as
$\mathbf{Z}^{(m)}=\{\mathbf{z}^{(m)}_i\}_{i=1}^{N_m}\in\mathbb{R}^{N_m\times d}$.

\subsection{Unified Multimodal Encoder Design}
\label{sec:encoder}

\begin{figure*}[htbp]
    \centering
    \vspace{-1em}
    \includegraphics[width=0.95\linewidth]{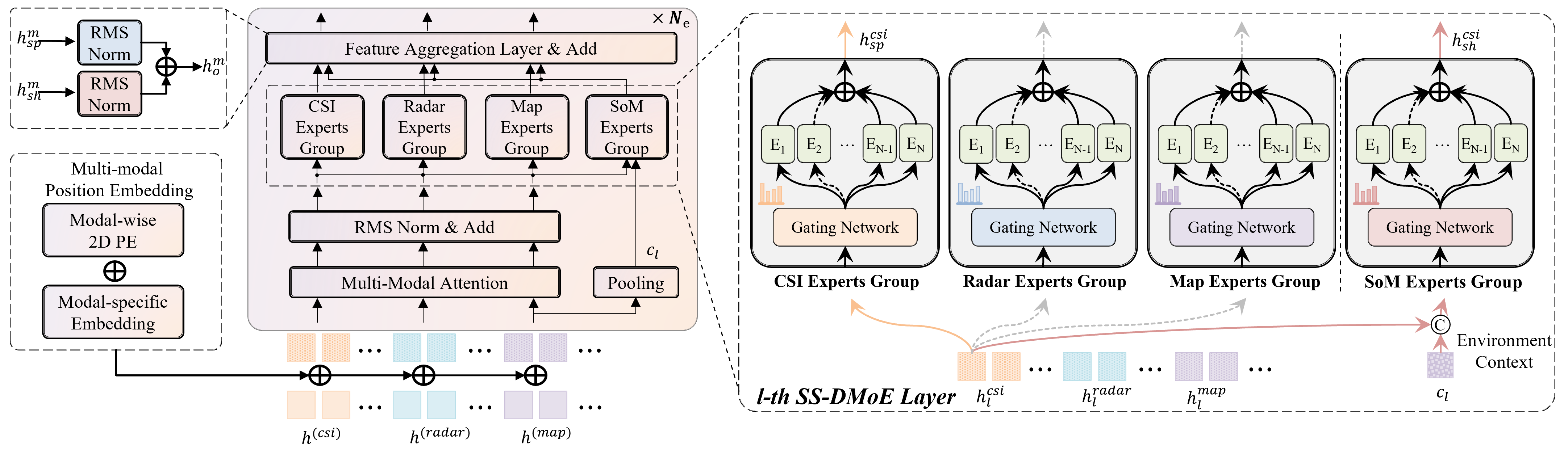}
    \caption{\RevisedText{An illustration of the unified multimodal encoder architecture in WiFo-MiSAC.}}
    \label{fig:network}
    \vspace{-1em}
    \end{figure*}

The encoder is responsible for uncovering cross-modal dependencies from multimodal token inputs, enabling modality alignment and fusion, and producing robust and informative joint representations.
It consists of {multimodal positional embeddings} and a stack of $N_e$ Transformer blocks.
Each transformer block comprises: (i) {modality-shared self-attention} over the concatenated token sequence (early fusion), and (ii) an {SS-DMoE feed-forward} layer with {shared/specific decoupling} and {context-aware sparse routing}.

\subsubsection{Multimodal Positional Embedding}
\label{sec:pos_embed_new}

Each token embedding is formed as
\begin{equation}
\mathbf{h}^{(m)}_i
=
\mathbf{z}^{(m)}_i
+
\mathbf{p}^{(m)}_i
+
\mathbf{u}^{(m)},
\label{eq:mm_embed_new}
\end{equation}
where $\mathbf{p}^{(m)}_i$ is a {modality-wise 2D absolute positional embedding}, and $\mathbf{u}^{(m)}$ is a {learnable modality identifier} shared by all tokens of modality $m$.
This design preserves within-modality geometry while disambiguating modalities under shared attention.

\subsubsection{Modality-shared Multimodal Attention}
\label{sec:mm_attention_new}

We concatenate token sequences from all modalities:
\begin{equation}
\mathbf{H}_\ell
=
\Big[
\mathbf{h}^{(\texttt{csi})}_{1:N_{\texttt{csi}}},
\mathbf{h}^{(\texttt{map})}_{1:N_{\texttt{map}}},
\mathbf{h}^{(\texttt{radar})}_{1:N_{\texttt{radar}}}
\Big]
\in\mathbb{R}^{N\times d},
\label{eq:concat_new}
\end{equation}
where $N=\sum_{m\in\mathcal{M}} N_m$.

At layer $\ell$, we first apply RMSNorm \cite{zhang2019root, bai2025qwen3} to the input hidden states and then perform shared multi-head self-attention (MSA) over \emph{all} tokens. The resulting attention output is added back through a residual connection:
\begin{equation}
\mathbf{H}'_\ell = \mathbf{H}_\ell + \mathrm{MSA}\!\left(\mathrm{RMSNorm}(\mathbf{H}_\ell)\right).
\label{eq:msa_new}
\end{equation}
This joint attention enables token-level cross-modal alignment and interaction, rather than relying on post-hoc fusion.
\subsubsection{SS-DMoE for Shared--Specific Cross-modal Fusion}
\label{sec:som_moe}

\RevisedText{Early fusion enables fine-grained interactions among heterogeneous modalities, but it may also introduce \emph{modality conflict}.
Different sensing modalities often reflect complementary yet partially inconsistent physical factors; enforcing them to pass through a single homogeneous transformation may suppress modality-specific cues or amplify spurious correlations.
Conversely, completely separating the modalities prevents the model from exploiting transferable cross-modal semantics.
To balance these two extremes, we introduce a shared--specific dynamic mixture-of-experts module, termed SS-DMoE, which explicitly decomposes the feed-forward transformation into a modality-shared pathway and a modality-specific pathway.}

Specifically, SS-DMoE maintains a shared expert pool
$\mathcal{E}^{(\texttt{som})}=\{E^{(\texttt{som})}_1,\ldots,E^{(\texttt{som})}_{K_s}\}$
for capturing modality-invariant patterns, and a set of modality-specific expert pools
$\mathcal{E}^{(m)}=\{E^{(m)}_1,\ldots,E^{(m)}_{K_m}\}$,
$m\in\mathcal{M}$, for preserving modality-dependent cues.
Each expert is implemented as a two-layer MLP with a GELU activation.
Given a token $\mathbf{x}$ from modality $m$, SS-DMoE computes sparse top-$k$ routing weights over the two expert pools as
\begin{subequations}
\begin{align}
\boldsymbol{\pi}^{(\texttt{som})}
&=
\mathrm{Softmax}_k\!\left(G^{(\texttt{som})}(\mathbf{x},\mathbf{c}_\ell)\right), \\
\boldsymbol{\pi}^{(m)}
&=
\mathrm{Softmax}_k\!\left(G^{(m)}(\mathbf{x})\right),
\label{eq:gating_new}
\end{align}
\end{subequations}
where $\mathrm{Softmax}_k(\cdot)$ denotes the \emph{top-$k$ truncated softmax}. Here, $\mathbf{c}_\ell$ is an environment context vector pooled from map tokens at layer $\ell$ (Sec.~\ref{sec:router}).
The shared and modality-specific expert outputs are then given by
\begin{subequations}
\begin{align}
\mathbf{y}^{(\texttt{som})}(\mathbf{x})
&=
\sum_{j\in\mathcal{T}^{(\texttt{som})}}
\pi^{(\texttt{som})}_j \, E^{(\texttt{som})}_j(\mathbf{x}), \\
\mathbf{y}^{(m)}(\mathbf{x})
&=
\sum_{j\in\mathcal{T}^{(m)}}
\pi^{(m)}_j \, E^{(m)}_j(\mathbf{x}),
\label{eq:moe_out_new}
\end{align}
\end{subequations}
where $\mathcal{T}^{(\texttt{som})}$ and $\mathcal{T}^{(m)}$ denote the selected top-$k$ expert indices from the shared and modality-specific pools, respectively.

As shown in Fig.~\ref{fig:network}, the SS-DMoE produces two parallel outputs for each token: a modality-shared branch from the SoM expert group and a modality-specific branch from the corresponding modality expert group. We then apply RMSNorm to each branch independently and fuse them with a simple element-wise sum, which keeps the two contributions on comparable scales while preserving their complementary roles. Finally, the fused expert output is added back to the token stream through a standard residual connection to form the block output.
\RevisedText{
The disentangling effect of SS-DMoE can be interpreted from the perspective of shared--private representation learning. Following the common principle of shared--private decomposition~\cite{meng2025br}, an effective multimodal representation should contain a shared component that captures modality-invariant semantics and a private component that preserves modality-dependent cues. SS-DMoE introduces this principle into the feed-forward transformation by explicitly factorizing the token-wise mapping as
\begin{equation}
\Phi_m(\mathbf{x},\mathbf{c}_\ell)
=
s(\mathbf{x},\mathbf{c}_\ell;\Theta_s)
+
p_m(\mathbf{x};\Theta_m),
\label{eq:ssdmoe_factorization}
\end{equation}
where $s(\cdot)$ is implemented by the shared SoM expert pool and $p_m(\cdot)$ is implemented by the expert pool dedicated to modality $m$. Therefore, instead of forcing all modalities to rely on a single homogeneous FFN, SS-DMoE provides two functionally different optimization pathways: one for cross-modal sharing and the other for modality-specific adaptation.}

\RevisedText{
From an optimization perspective, a fully shared feed-forward module entangles the learning signals of different modalities within the same parameter space. During back-propagation, gradients from all modalities are accumulated on the same FFN parameters. Since heterogeneous modalities may reflect complementary but partially inconsistent physical factors, their gradients can favor different update directions. Such coupled optimization may cause the shared parameters to overfit modality-specific variations, suppress useful private cues, or even introduce negative transfer, which constitutes an optimization-level source of modality conflict in early fusion.
}

\RevisedText{
SS-DMoE mitigates this issue by explicitly separating the gradient flow into shared and modality-specific pathways. The shared SoM experts receive optimization signals from all modalities and are thus encouraged to capture transformations that are consistently useful across modalities, such as modality-invariant environmental semantics. In contrast, each modality-specific expert pool is updated only by tokens from its corresponding modality, allowing it to model residual variations induced by sensor-specific noise, resolution, geometry, or measurement characteristics. Therefore, sharing-oriented and specialization-oriented updates are not forced to compete entirely within the same FFN parameters.
}

\RevisedText{
The sparse top-$k$ routing mechanism further regularizes this shared--specific design. For each token, only a small subset of shared and modality-specific experts is activated, which reduces unnecessary parameter co-adaptation and encourages different experts to specialize in distinct transformation patterns. In addition, the shared router is conditioned on the environment context $\mathbf{c}_\ell$, allowing the shared SoM pathway to adapt cross-modal semantics according to the surrounding scene, while the modality-specific pathway preserves local modality-dependent information from the token itself.
}

\RevisedText{
Overall, SS-DMoE provides an architectural and optimization-level inductive bias for shared--private representation learning. It does not assume that modality-shared and modality-specific factors can be perfectly separated; instead, it encourages the model to balance transferable cross-modal semantics and residual modality-dependent cues through two complementary pathways. Deriving formal identifiability guarantees for shared--private multimodal decomposition remains an interesting direction for future work.}

\subsubsection{Environment Context Embedding for Scene-adaptive Routing}
\label{sec:router}

Wireless modalities such as CSI and radar individually exhibit regularities that can often be captured without explicit geometric priors. However, {multimodal fusion} introduces an additional challenge: the {alignment/mapping mechanism across modalities} is strongly scene-dependent, because the environment modulates how different sensing channels co-vary (e.g., through geometry-dependent multipath and occlusion patterns). Consequently, using a fixed fusion pathway to extract {modality-shared} factors can be suboptimal under scene shifts, even if each modality alone remains learnable.

To address this, we condition the routing of the {shared expert pool} (SoM experts), which is responsible for extracting {modality-shared} representations, on an explicit {environment context} derived from map tokens. In contrast, the routing for {modality-specific} experts does {not} use map context, since modality-exclusive cues are expected to follow more stable, modality-intrinsic statistics across scenes.

\paragraph{Context pooling from map tokens}
Rather than feeding dense map tokens directly into the router, we summarize them into a compact scene descriptor to provide a stable routing signal and avoid redundancy.
Let $\mathbf{H}^{(\texttt{map})}_\ell=\{\mathbf{h}^{(\texttt{map})}_{i,\ell}\}_{i=1}^{N_{\texttt{map}}}$ denote the map token states at layer $\ell$.
We compute a mean-pooled context vector:
\begin{equation}
\mathbf{c}_\ell
=
\frac{1}{N_{\texttt{map}}}
\sum_{i=1}^{N_{\texttt{map}}}
\mathbf{h}^{(\texttt{map})}_{i,\ell}.
\label{eq:context_pool_new}
\end{equation}

\paragraph{Asymmetric context-aware gating}
For the \emph{shared} router, we inject $\mathbf{c}_\ell$ by concatenation:
\begin{equation}
G^{(\texttt{som})}(\mathbf{x},\mathbf{c}_\ell)
=
\mathbf{W}_g^{(\texttt{som})}\,[\mathbf{x};\mathbf{c}_\ell]+\mathbf{b}_g^{(\texttt{som})},
\label{eq:context_router_som}
\end{equation}
so that expert selection adapts to the current scene and can realize scene-conditioned cross-modal correspondence.
For the \emph{modality-specific} routers, we intentionally \emph{exclude} map context:
\begin{equation}
G^{(m)}(\mathbf{x})
=
\mathbf{W}_g^{(m)}\,\mathbf{x}+\mathbf{b}_g^{(m)},
\label{eq:context_router_spec}
\end{equation}
which preserves stable, modality-intrinsic specialization without over-conditioning on geometry.

This asymmetric design makes the shared pathway explicitly scene-adaptive, facilitating the learning of modality-shared factors for cross-modal fusion, while keeping modality-specific routing scene-agnostic to preserve stable, modality-intrinsic specialization and improve robustness under distribution shifts. Notably, when map data is unavailable, we set $\mathbf{c}_\ell$ to the zero vector, reducing the shared router to an $\mathbf{x}$-only form and enabling inference without architectural changes.

\subsection{Hybrid Pre-training Strategy}
\label{sec:pretrain}

We pre-train WiFo-MiSAC with a {hybrid objective} that jointly optimizes {cross-modal masked reconstruction} and {multimodal contrastive learning}, together with a standard {MoE load-balancing} regularizer. The masked reconstruction task encourages the encoder to form holistic multimodal representations that support cross-modal completion under partial observations, while contrastive learning explicitly promotes {shared} cross-modal factors, strengthening mutual benefit across modalities. The overall objective is
\begin{equation}
\mathcal{L}_{\mathrm{pre}}
=
\mathcal{L}_{\mathrm{mask}}
+\lambda_{\mathrm{cl}}\mathcal{L}_{\mathrm{cl}}
+\lambda_{\mathrm{lb}}\mathcal{L}_{\mathrm{lb}}.
\label{eq:pretrain_total_new}
\end{equation}

\subsubsection{Cross-modal Masked Reconstruction}
\label{sec:mask_modeling_new}

\paragraph{MAE-style masking}
Let $\mathbf{H}^{(m)}\in\mathbb{R}^{N_m\times d}$ be the token sequence of modality $m$.
Following MAE-style training, we sample a visible index set $\mathcal{V}^{(m)}$ (equivalently, a masked set $\mathcal{S}^{(m)}$) and feed \emph{only} the visible tokens $\mathbf{H}^{(m)}_{\mathcal{V}^{(m)}}$ into the unified encoder.
After encoding, we append learnable mask tokens at the decoder input to represent the missing positions, and apply a lightweight modality-specific decoder $\mathcal{D}^{(m)}$ to predict the full signal:
\begin{equation}
\widehat{\mathbf{X}}^{(m)}
=
\mathcal{D}^{(m)}\!\left(
\left[\mathrm{Enc}\!\left(\widetilde{\mathbf{H}}\right);\,
\mathbf{E}_{\texttt{mask}}^{(m)}
\right]
\right),
\label{eq:mae_decode_new}
\end{equation}
where $\widetilde{\mathbf{H}}=[\mathbf{H}^{(\texttt{csi})}_{\mathcal{V}^{(\texttt{csi})}};\mathbf{H}^{(\texttt{map})}_{\mathcal{V}^{(\texttt{map})}};\mathbf{H}^{(\texttt{radar})}_{\mathcal{V}^{(\texttt{radar})}}]$ denotes the concatenated visible tokens, and $\mathbf{E}_{\texttt{mask}}^{(m)}$ denotes the set of learnable mask tokens inserted to fill the missing slots of modality $m$ at the decoder side. This design reduces encoder computation and forces the encoder to model cross-modal dependencies from partial observations.

\paragraph{CSI-centric masking schemes}
We adopt three complementary masking patterns, namely \textit{Random Masking}, \textit{Frequency-domain Masking}, and \textit{Comb Masking}. 
First, \textit{Random Masking} is applied to all modalities, where the masking ratio is uniformly sampled as $r_{\mathrm{rand}}\sim\mathcal{U}(0.1,0.5)$. This strategy improves robustness against arbitrary missing observations. 
Second, for CSI, we introduce \textit{Frequency-Structured Masking}, which masks contiguous stripes on the 2D CSI token grid along the frequency dimension. The masking ratio is similarly sampled as $r_{\mathrm{freq}}\sim\mathcal{U}(0.1,0.5)$, encouraging the model to recover structured spectral correlations. 
Third, we employ \textit{Comb Masking} for CSI, where visible tokens are sparsely retained in a comb-style pattern with a uniformly sampled frequency-domain spacing $N_s\sim\mathcal{U}\{4,16\}$, while the remaining CSI tokens are masked. This design mimics comb-type pilot sampling in practical OFDM systems and promotes the learning of interpolation-aware representations.


We emphasize CSI because, in many communication settings, CSI is frequently \emph{partially observed} (e.g., sparse pilots, subcarrier dropping, or bandwidth/time subsampling), and the ability to reconstruct CSI under diverse missing patterns directly supports a broad range of downstream tasks. In contrast, radar and map inputs are typically available in a more complete form and are less subject to protocol-driven sparsification. Accordingly, we diversify CSI masking to expose the model to realistic partial-observation regimes, improving the robustness of CSI reconstruction and strengthening its role as a bridge modality for multimodal fusion.
\paragraph{Reconstruction objective}
The reconstruction loss is computed only on the masked support:
\begin{equation}
\mathcal{L}_{\mathrm{mask}}
=
\sum_{m\in\mathcal{M}}
\left\|
\Omega^{(m)} \odot \left(\widehat{\mathbf{X}}^{(m)}-\mathbf{X}^{(m)}\right)
\right\|^{2},
\label{eq:mask_loss_new}
\end{equation}
where $\Omega^{(m)}$ is the binary mask indicating the masked positions under the corresponding scheme.

\begin{figure*}[htbp]
    \centering
    \vspace{-1em}
    \includegraphics[width=0.9\linewidth]{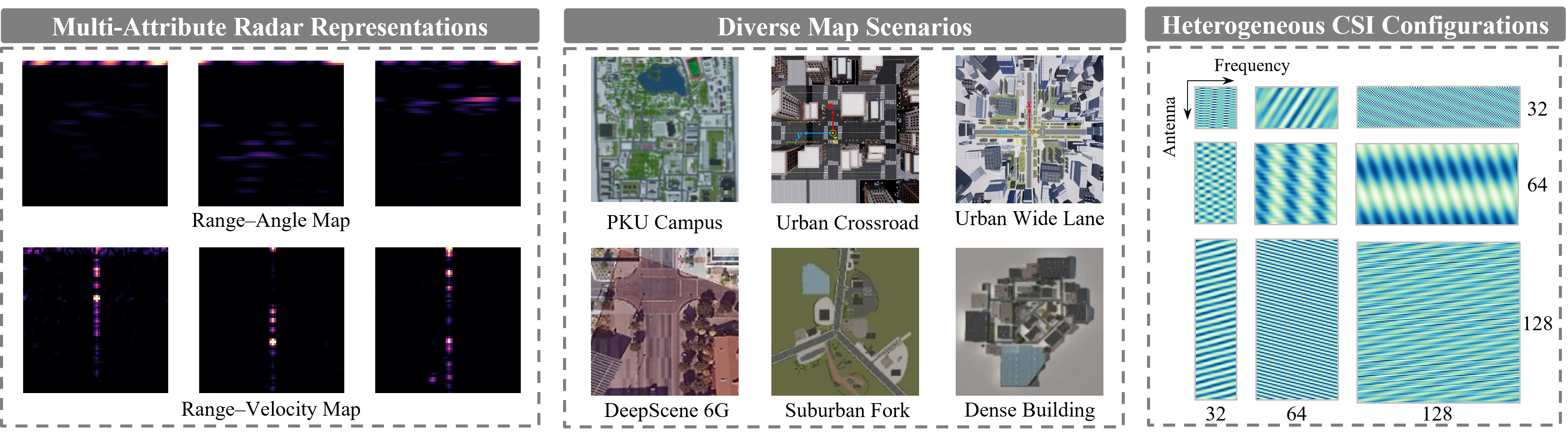}
    \caption{\RevisedText{Visualization of the constructed pre-training dataset.}}
    \label{pic:dataset}
    \vspace{-1em}
    \end{figure*}

\subsubsection{CSI-Anchored Cross-modal Contrastive Alignment}
\label{sec:contrastive_new}

Masked modeling promotes information propagation but does not explicitly enforce a shared semantic space. We therefore introduce a {CSI-anchored} contrastive objective that aligns map/radar representations to CSI. Concretely, we treat the temporally synchronized triplet $(\texttt{CSI}, \texttt{Radar}, \texttt{Map})$ from the same scene as a \emph{positive} association, and construct negatives from other scenes within the mini-batch. Importantly, this contrastive loss is computed \emph{only} for samples where all three modalities are available, ensuring that the alignment signal is grounded on complete cross-modal evidence. For scenes with missing modalities, we simply omit the contrastive term and rely on masked modeling objectives to train the available modalities, avoiding noisy or ill-defined supervision under modality absence.

\paragraph{Projection and pooling}
For each modality $m$, we apply a projection head and average-pool tokens:
\begin{subequations}
\begin{align}
&\mathbf{U}^{(m)} = \mathbf{H}^{(m)}\mathbf{W}_{\mathrm{p}}^{(m)} + \mathbf{b}_{\mathrm{p}}^{(m)},\qquad \\
&\widetilde{\mathbf{z}}^{(m)} = \frac{1}{N_m}\sum_{i=1}^{N_m}\mathbf{U}^{(m)}_i,\qquad  
\label{eq:contrastive_pool_new}
\end{align}
\end{subequations}

\paragraph{InfoNCE with CSI as anchor}
Given a minibatch of size $B$, $(\widetilde{\mathbf{z}}^{(\texttt{csi})}_i,\widetilde{\mathbf{z}}^{(m)}_i)$ are positives for $m\in\{\texttt{map},\texttt{radar}\}$:
\begin{subequations}
\begin{align}
\mathcal{L}_{\mathrm{cl}}^{(m)}
=
-\frac{1}{B}\sum_{i=1}^{B}&
\log
\frac{\exp\left((\widetilde{\mathbf{z}}^{(\texttt{csi})}_i)^\top \widetilde{\mathbf{z}}^{(m)}_i/\tau\right)}
{\sum_{j=1}^{B}\exp\left((\widetilde{\mathbf{z}}^{(\texttt{csi})}_i)^\top \widetilde{\mathbf{z}}^{(m)}_j/\tau\right)}, \\
&\mathcal{L}_{\mathrm{cl}} = \mathcal{L}_{\mathrm{cl}}^{\texttt{radar}} + \mathcal{L}_{\mathrm{cl}}^{\texttt{map}},
\label{eq:infonce_new}
\end{align}
\end{subequations}
where $\tau$ is a temperature hyper-parameter.

\subsubsection{MoE Load Balancing}
\label{sec:load_balance_new}

To stabilize sparse routing and prevent expert collapse, we apply a standard load-balancing regularizer to each expert pool \cite{lepikhin2020gshard,fedus2022switch}. Let $p_{i,e}$ be the routing weight from token $i$ to expert $e$, and $\mathbb{I}_{i,e}$ indicate whether expert $e$ is selected in top-$k$. Define expert \emph{importance} and \emph{load}:
\begin{equation}
\mathrm{Imp}(e)=\frac{1}{N}\sum_{i=1}^{N}p_{i,e},
\qquad
\mathrm{Load}(e)=\frac{1}{N}\sum_{i=1}^{N}\mathbb{I}_{i,e},
\label{eq:imp_load_new}
\end{equation}
and penalize skewed utilization by
\begin{equation}
\mathcal{L}_{\mathrm{lb}}
=
\frac{1}{K}\sum_{e=1}^{K}\mathrm{Imp}(e)\,\mathrm{Load}(e),
\label{eq:lb_new}
\end{equation}
where $K$ is the number of experts in the pool and $N$ is the number of routed tokens.


\section{Numerical Simulations}
In this section, we first present the details of the constructed pre-training datasets. We then outline the simulation setup in detail. Finally, we comprehensively evaluate and analyze the performance of WiFo-MiSAC. 
\subsection{Pre-training Dataset}

\begin{table*}[t]
    \centering
    \vspace{-1em}
\caption{Dataset configurations and data splits.}
    \label{tab:dataset_config}
    \setlength{\tabcolsep}{3pt}
    \renewcommand{\arraystretch}{1}
    \resizebox{\textwidth}{!}{%
    \begin{tabular}{llccccccccc}
        \toprule
        Data Source &
        Scenario &
        Dataset ID &
        Fre. (GHz) &
        Num Subc. &
        $BW$ (MHz) &
        BS Ant. &
        Samples &
        CSI/Radar/Map &
        \makecell{Pre-train\\set} &
        \makecell{Test \\ set} \\
        \midrule

        \multirow{2}{*}{M$^{3}$SC \cite{cheng2023m}} &
        \multirow{2}{*}{Urban Crossroad} &
        CF1--CF9 & 28.0 & 64--128 & 20--80   & 64--128 & 13,500 &
        \multirow{2}{*}{$\checkmark/\checkmark/\checkmark$} &
        \multirow{2}{*}{1--4} &
        \multirow{2}{*}{5--9} \\
        & &
        CC1--CC9 & 28.0 & 64--128 & 200--800 & 64--128 & 13,500 &
        & & \\
        \midrule

        \multirow{8}{*}{SynthSoM \cite{cheng2025synthsom}} &
        \multirow{2}{*}{Suburban Fork} &
        FF1--FF16 & 28.0 & 64--128 & 20--80   & 64--128 & 22,500 &
        \multirow{2}{*}{$\checkmark/\checkmark/\checkmark$} &
        \multirow{2}{*}{1--14} &
        \multirow{2}{*}{15--16} \\
        & &
        FC1--FC16 & 28.0 & 64--128 & 200--800 & 64--128 & 22,500 &
        & & \\
        \cmidrule(lr){2-11}

        & \multirow{2}{*}{PKU Campus} &
        PF1--PF15 & 5.9  & 32--128 & 2--5     & 32--128 & 67,500 &
        \multirow{2}{*}{$\checkmark/\checkmark/\checkmark$} &
        \multirow{2}{*}{1--9} &
        \multirow{2}{*}{10--15} \\
        & &
        PC1--PC15 & 5.9  & 32--128 & 20--50   & 32--128 & 67,500 &
        & & \\
        \cmidrule(lr){2-11}

        & \multirow{2}{*}{Dense Building} &
        DF1--DF6 & 4.95 & 32--128 & 1--8     & 64--128 & 18,000 &
        \multirow{2}{*}{$\checkmark/-/\checkmark$} &
        \multirow{2}{*}{1--4} &
        \multirow{2}{*}{5--6} \\
        & &
        DC1--DC6 & 4.95 & 32--128 & 10--80   & 64--128 & 18,000 &
        & & \\
        \cmidrule(lr){2-11}

        & \multirow{2}{*}{Urban Wide Lane} &
        WF1--WF25 & 28.0 & 64--128 & 20--80   & 64--128 & 22,500 &
        \multirow{2}{*}{$\checkmark/-/\checkmark$} &
        \multirow{2}{*}{1--10} &
        \multirow{2}{*}{11--25} \\
        & &
        WC1--WC25 & 28.0 & 64--128 & 200--800 & 64--128 & 22,500 &
        & & \\
        \midrule

        DeepSense 6G \cite{alkhateeb2023deepsense} &
        Scenario 30--35 &
        S1--S6 &
        \multicolumn{4}{c}{\textemdash} &
        30,000 &
        $-/\checkmark/-$ &
        1--3 &
        4--6 \\
        \bottomrule
    \end{tabular}%
    }
\begin{tablenotes}
\footnotesize
\item Note: ``CSI/Radar/Map'' indicates the availability of each modality, where $\checkmark$ denotes available and ``--'' denotes missing.
\end{tablenotes}
\vspace{-1em}
\end{table*}

We construct a large-scale multimodal pre-training dataset by consolidating CSI, radar, and map data from M$^{3}$SC~\cite{cheng2023m}, SynthSoM~\cite{cheng2025synthsom}, and DeepSense-6G~\cite{alkhateeb2023deepsense}.
Overall, the dataset contains over 1B complex CSI entries (counted over antennas, and subcarriers) and more than 200k time-synchronized CSI--radar--map triplets. 
\RevisedText{Fig.~\ref{pic:dataset} provides representative examples of the collected modalities, highlighting the diversity of radar signatures, deployment scenarios, and CSI configurations covered by the dataset, which is critical for pre-training a foundation model with robust and transferable wireless representations.}
\RevisedText{Specifically,} CSI is collected across diverse environments, and spans a wide range of carrier frequencies (4.95/5.9/28~GHz), subcarrier counts (32--128), bandwidths (1--800~MHz), and BS array sizes (32--128), yielding substantial variability in propagation and spatial sampling conditions.
All radar measurements are converted into $64\times64$ range--velocity and range--angle representations, while all maps are resized to $256\times256$ resolution.
Table~\ref{tab:dataset_config} summarizes the per-scenario configurations and splits.
Notably, for each scenario we provide two complementary CSI variants: a narrowband \emph{fine-grained} setting and a wideband \emph{coarse-grained} setting, which exposes the model to multi-resolution spectral observations and encourages scale-robust channel representations that can flexibly adapt to heterogeneous bandwidth regimes.
Dataset IDs follow a structured convention: in an ID such as ``PF1'', the first character denotes the scenario, the second character indicates CSI granularity (F: fine-grained, C: coarse-grained), and the trailing number indexes the communication link within that scenario.
We split each scenario into pre-training and disjoint OOD test subsets by link indices, ensuring that evaluation is conducted on unseen links, \RevisedText{each corresponding to a new BS--UE geometric realization, user trajectory, and local map instance.}

We also remark that the symbol ``--'' in Table~\ref{tab:dataset_config} indicates missing modalities.
Specifically, \textit{Dense Building} and \textit{Urban Wide Lane} do not provide radar measurements, whereas \textit{DeepSense-6G} only contains radar data.
We nevertheless retain these partially-observed samples in the pre-training dataset to reflect a realistic deployment condition, where fully paired multimodal acquisitions are often scarce, and the majority of data is inherently unpaired.
A key advantage of our framework over conventional contrastive learning pipelines is its ability to leverage such unpaired multimodal data without discarding samples or forcing unreliable pseudo-pairing, thereby continuously improving representation learning during pre-training and strengthening the extracted multimodal features.

\subsection{Simulation Setup}

\subsubsection{Downstream Tasks}
\label{sec:downstream_tasks_baselines}
We consider five representative downstream tasks spanning wireless communication and sensing to comprehensively assess the transferability of the proposed pre-trained general-purpose multimodal representation across heterogeneous application objectives. For each task, we evaluate both unimodal inputs and multimodal inputs to demonstrate that WiFo-MiSAC remains effective under different modality availability, rather than being limited to multimodal inference as in most task-specific multimodal methods.

\begin{itemize}
    \item \textbf{Frequency-domain Channel Prediction \cite{liu2024llm4cp}.}
    Accurate CSI is fundamental to modern wireless systems, as it directly impacts key physical-layer procedures such as precoding, adaptive modulation and coding, and power control.
    In this task, the model takes as input the CSI over $N_h$ consecutive subcarriers (and, when available, synchronous auxiliary sensing) and predicts the CSI over the subsequent $N_p$ subcarriers.
    This setting evaluates whether the learned representation captures stable propagation- and geometry-related factors and can extrapolate channel variations in the frequency domain.

    \item \textbf{Channel Estimation \cite{liu2025llm4wm}.}
    Channel estimation can be formulated as frequency-domain interpolation and denoising from sparse pilot observations. Specifically, pilots are inserted every $N_s$ subcarriers, so the model input is the CSI measured only at pilot subcarriers (optionally augmented with synchronous auxiliary sensing), and the output is the reconstructed full-resolution CSI over all subcarriers.
    This task evaluates the model’s ability to leverage generic channel structures, including multipath-induced frequency correlation and sparsity, to recover accurate CSI from incomplete and noisy measurements.
    
    \item \textbf{mmWave Beam Prediction.}
    Millimeter-wave (mmWave) bands are critical for 5G and emerging 6G systems, but exhaustive beam sweeping incurs substantial overhead.
    Since practical devices often support both sub-6\,GHz and mmWave front-ends and these modalities observe the same geometry, sub-6\,GHz CSI or radar measurements can provide informative priors for mmWave beam alignment.
    In this task, the input is an instantaneous subcarrier-by-antenna CSI sample or a radar sample (with multimodal variants using both), and the output is the codebook index of the optimal mmWave beam.
    This task evaluates the model's capability for cross-modality and cross-band geometry-aware alignment and robust beam selection.

    \item \textbf{Distance Estimation.}
    Distance estimation aims to infer the transmitter--receiver range from wireless observations, enabling ranging and localization services in environments where GNSS is degraded or unavailable.
    The input is a CSI sample at a given time instant (optionally augmented with auxiliary sensing), and the output is a scalar distance.
    This task probes the representation's ability to extract delay-related cues and remain robust under multipath and cluttered propagation.

    \item \textbf{AoA Estimation.}
    Angle-of-arrival (AoA) estimation infers the dominant-path arrival angle from CSI and provides critical priors for downstream beamforming and localization.
    The input is a CSI sample at a given time instant (optionally with auxiliary sensing), and the output is the AoA of the channel's dominant path.
    This task assesses spatial-spectrum super-resolution, dominant-path separability, and robust generalization in rich multipath environments.
\end{itemize}

\subsubsection{Baselines}
\label{sec:baselines}
We compare WiFo-MiSAC with representative baselines from task-specific unimodal learning, task-specific multimodal fusion, multimodal foundation models, and wireless foundation models. For all methods, we follow the original training recipes whenever possible and only modify the input tokenizers and output heads to match our sensing modalities and downstream tasks.

\begin{itemize}
\item \textbf{Task-specific unimodal baselines.}
We include widely used unimodal methods for wireless sensing and communication tasks, including Transformer~\cite{jiang2022accurate} for channel prediction, Channelformer~\cite{luan2023channelformer} for pilot-based channel estimation, LLM4CP~\cite{liu2024llm4cp} and its channel-estimation variant for LLM-based CSI modeling, BP-DNN~\cite{alrabeiah2020deep} for beam prediction, and WiT~\cite{salihu2022attention} for distance and AoA estimation.

\item \RevisedText{\textbf{Task-specific multimodal fusion baselines.}
We compare with representative non-foundation multimodal fusion schemes, including Trans-MM~\cite{nam2025multi}, CNN-MM~\cite{zhang2024integrated}, MSTN~\cite{song2022multimodal}, and SelectFusion~\cite{chen2022learning}. These methods cover cross-attention fusion, late CNN fusion, multi-stage/multi-stream feature interaction, and reliability-aware selective sensor fusion. We adapt their fusion modules to CSI, radar, and map tokens, while keeping their task-specific training paradigm unchanged.}

\item \RevisedText{\textbf{Multimodal foundation model baselines.}
To compare with recent MoE-based multimodal foundation models, we include EVE~\cite{chen2024eve} and BR-MoE~\cite{meng2025br}. EVE introduces modality-aware sparse MoE into a shared multimodal Transformer, while BR-MoE uses route-dynamic experts to model modality-agnostic and modality-specific information. We adapt them to our wireless modalities and evaluate them under the same pre-training and downstream protocols, with their backbone capacity and key architectural settings matched to WiFo-MiSAC to ensure a fair comparison focused on routing and expert-decomposition designs.}

\item \textbf{Wireless foundation model baseline.}
We further include the WiFo-based scheme~\cite{liu2024wifo} as a foundation-model control baseline. It shares the same backbone and downstream adaptation protocol as WiFo-MiSAC, but is pre-trained using unimodal wireless data only. This comparison isolates the benefit of multimodal pre-training and evaluates whether cross-modal pre-training improves both unimodal and multimodal inference.

\end{itemize}

\subsubsection{Network and Pre-training Settings}
We pre-train WiFo-MiSAC using an MoE-based encoder--decoder architecture. 
The backbone contains a 4-layer encoder and a 2-layer decoder, both with a feature dimension of 512 and 8 attention heads. 
Each expert group consists of 8 experts, among which 4 are activated for each forward pass. 
For tokenization, CSI samples are partitioned with a patch size of $(4,4)$, whereas radar and map inputs use a patch size of $(8,8)$. 
Pre-training is performed on a server equipped with four Intel Xeon Platinum 8358P CPUs, four NVIDIA RTX 5090 GPUs, and 188~GB RAM.

To enhance robustness against heterogeneous signal qualities, additive noise is injected into CSI during pre-training, where the SNR is uniformly sampled from 10 to 25~dB. 
All datasets listed in Table~\ref{tab:dataset_config} are converted into the required input format, shuffled, and sequentially fed into the model for parameter updates. 
Unless otherwise specified, the model is pre-trained for 150 epochs and fine-tuned for 100 epochs with a batch size of 64. 
We use the Adam optimizer with $\beta_1=0.9$ and $\beta_2=0.999$, together with a cosine annealing learning-rate schedule over 150 epochs, where the learning rate is bounded between $1\times10^{-6}$ and $1\times10^{-4}$. 
For loss re-weighting, we set $\lambda_{\mathrm{cl}}=5\times10^{-4}$ and $\lambda_{\mathrm{lb}}=5\times10^{-2}$, and use a temperature $\tau=0.07$ in the contrastive objective.


\subsection{Performance Evaluation}

\begin{table*}[htbp]
\vspace{-1em}
\centering
\caption{\RevisedText{Channel prediction NMSE (dB) of the proposed method and baselines on ID/OOD splits with unimodal and multimodal inputs. Best and second-best results are highlighted by \best{bold} and \second{underline}, respectively. ``\textemdash'' indicates missing modalities where multimodal inputs are unavailable. $^\dagger$ indicates zero-shot OOD inference without fine-tuning.}}
\label{tab:CP_results}
\setlength{\tabcolsep}{1pt}
\renewcommand{\arraystretch}{1.15}
\resizebox{\textwidth}{!}{
\begin{tabular}{c c c c c c c c c c c c c c c c}
\toprule
\multirow{2}{*}{\makecell{\textbf{Evaluation}\\\textbf{split}}} &
\multirow{2}{*}{\textbf{Dataset ID}} &
\multicolumn{6}{c}{\textbf{Unimodal input (CSI-only)}} &
\multicolumn{8}{c}{\textbf{Multimodal input (CSI \& Radar \& Map)}} \\
\cmidrule(lr){3-8}\cmidrule(lr){9-16}
& &
\textbf{WiFo-MiSAC$^\dagger$} & \textbf{EVE$^\dagger$} & \textbf{BR-MoE$^\dagger$} & \textbf{WiFo-based$^\dagger$} & \textbf{LLM-based} & \textbf{Transformer} &
\textbf{WiFo-MiSAC$^\dagger$} & \textbf{EVE$^\dagger$} & \textbf{BR-MoE$^\dagger$} & \textbf{WiFo-based} & \textbf{MSTN} & \textbf{SelectFusion} & \textbf{Trans-MM} & \textbf{CNN-MM} \\
\midrule

\multirow{6}{*}{\makecell{In-distribution\\(ID)}} & CC1--CC4 &
\best{-18.220} & -14.898 & \second{-15.227} & -11.525 & -9.109 & -5.912 &
\best{-18.593} & -15.444 & \second{-15.896} & -11.873 & -10.327 & -13.936 & -10.051 & -13.699 \\
& FC1--FC14 &
\best{-18.547} & -14.754 & \second{-15.775} & -10.487 & -1.888 & -1.304 &
\best{-18.715} & -15.163 & \second{-16.204} & -10.743 & -6.481 & -9.177 & -10.176 & -5.815 \\
& DC1--DC4 &
\best{-18.546} & -16.059 & \second{-16.316} & -13.420 & -8.592 & -6.723 &
\textemdash & \textemdash & \textemdash & \textemdash & \textemdash & \textemdash & \textemdash & \textemdash \\
& PC1--PC9 &
\best{-19.036} & -16.605 & \second{-16.782} & -16.429 & -12.071 & -11.501 &
\best{-19.627} & -17.502 & \second{-18.233} & -17.143 & -11.944 & -14.587 & -14.053 & -15.692 \\
& WC1--WC10 &
\best{-16.432} & -11.086 & \second{-11.993} & -4.717 & -1.243 & -0.613 &
\textemdash & \textemdash & \textemdash & \textemdash & \textemdash & \textemdash & \textemdash & \textemdash \\
& \textbf{Avg.} &
\best{-18.156} & -14.680 & \second{-15.218} & -11.316 & -6.581 & -5.211 &
\best{-18.978} & -16.036 & \second{-16.778} & -13.253 & -9.584 & -12.566 & -11.427 & -11.735 \\
\midrule

\multirow{6}{*}{\makecell{Out-of-distribution\\(OOD)}} & CC5 &
\best{-18.546} & \second{-15.690} & -15.435 & -10.939 & -10.372 & -9.438 &
\best{-18.878} & -15.996 & -15.723 & -12.279 & -16.087 & \second{-16.742} & -10.000 & -11.229 \\
& FC15--FC16 &
\best{-19.065} & -15.442 & \second{-16.786} & -11.534 & -1.093 & -0.838 &
\best{-19.161} & -15.861 & \second{-17.213} & -12.202 & -6.273 & -8.579 & -7.027 & -5.740 \\
& DC5--DC6 &
\best{-18.456} & \second{-16.234} & -16.226 & -13.214 & -10.098 & -8.942 &
\textemdash & \textemdash & \textemdash & \textemdash & \textemdash & \textemdash & \textemdash & \textemdash \\
& PC10--PC15 &
\best{-18.188} & -13.915 & \second{-14.506} & -10.405 & -8.932 & -7.628 &
\best{-19.176} & -15.277 & \second{-16.180} & -13.373 & -9.817 & -11.355 & -8.369 & -11.060 \\
& WC11--WC25 &
\best{-16.658} & -9.567 & \second{-10.917} & -3.190 & -1.410 & -0.986 &
\textemdash & \textemdash & \textemdash & \textemdash & \textemdash & \textemdash & \textemdash & \textemdash \\
& \textbf{Avg.} &
\best{-18.183} & -14.169 & \second{-14.774} & -9.856 & -6.381 & -5.567 &
\best{-19.072} & -15.711 & \second{-16.372} & -12.618 & -10.726 & -12.225 & -8.465 & -9.343 \\
\bottomrule
\end{tabular}}
\vspace{-1em}
\end{table*}

\begin{table*}[htbp]
\vspace{-1em}
\centering
\caption{\RevisedText{Channel estimation NMSE (dB) of the proposed method and baselines on ID/OOD splits with unimodal and multimodal inputs. Best and second-best results are highlighted by \best{bold} and \second{underline}, respectively. ``\textemdash'' indicates missing modalities where multimodal inputs are unavailable. $^\dagger$ indicates zero-shot OOD inference without fine-tuning.}}
\label{tab:CE_results}
\setlength{\tabcolsep}{1pt}
\renewcommand{\arraystretch}{1.15}
\resizebox{\textwidth}{!}{
\begin{tabular}{c c c c c c c c c c c c c c c c}
\toprule
\multirow{2}{*}{\makecell{\textbf{Evaluation}\\\textbf{split}}} &
\multirow{2}{*}{\textbf{Dataset ID}} &
\multicolumn{6}{c}{\textbf{Unimodal input (CSI-only)}} &
\multicolumn{8}{c}{\textbf{Multimodal input (CSI \& Radar \& Map)}} \\
\cmidrule(lr){3-8}\cmidrule(lr){9-16}
& &
\textbf{WiFo-MiSAC$^\dagger$} & \textbf{EVE$^\dagger$} & \textbf{BR-MoE$^\dagger$} & \textbf{WiFo-based$^\dagger$} & \textbf{LLM-based} & \textbf{Channelformer} &
\textbf{WiFo-MiSAC$^\dagger$} & \textbf{EVE$^\dagger$} & \textbf{BR-MoE$^\dagger$} & \textbf{WiFo-based} & \textbf{MSTN} & \textbf{SelectFusion} & \textbf{Trans-MM} & \textbf{CNN-MM} \\
\midrule

\multirow{6}{*}{\makecell{In-distribution\\(ID)}} & CF1--CF4 &
\best{-22.634} & -19.278 & -15.220 & \second{-21.850} & -17.584 & -17.991 &
\best{-23.726} & -20.288 & -21.308 & \second{-23.682} & -18.008 & -18.153 & -21.516 & -21.901 \\
& FF1--FF14 &
\best{-21.331} & -18.342 & -16.766 & \second{-20.180} & -16.475 & -14.437 &
\best{-23.146} & -19.509 & -19.271 & \second{-22.142} & -20.006 & -21.108 & -19.055 & -20.852 \\
& DF1--DF4 &
\best{-22.164} & -18.684 & -18.069 & \second{-19.166} & -14.545 & -15.602 &
\textemdash & \textemdash & \textemdash & \textemdash & \textemdash & \textemdash & \textemdash & \textemdash \\
& PF1--PF9 &
\best{-19.296} & -9.321 & -8.468 & \second{-17.773} & -14.819 & -16.951 &
\best{-22.190} & -16.865 & -18.596 & -19.282 & -20.248 & -19.161 & -18.073 & \second{-20.826} \\
& WF1--WF10 &
\best{-19.602} & -13.163 & -13.373 & -12.983 & -13.052 & \second{-14.005} &
\textemdash & \textemdash & \textemdash & \textemdash & \textemdash & \textemdash & \textemdash & \textemdash \\
& \textbf{Avg.} &
\best{-21.005} & -15.758 & -14.379 & \second{-18.390} & -15.295 & -15.797 &
\best{-22.787} & -18.887 & -19.725 & \second{-21.702} & -19.421 & -19.474 & -19.548 & -21.193 \\
\midrule

\multirow{6}{*}{\makecell{Out-of-distribution\\(OOD)}} & CF5 &
\best{-22.572} & -19.952 & -15.266 & \second{-21.599} & -18.117 & -19.165 &
\best{-23.708} & -20.813 & -21.106 & \second{-23.668} & -15.500 & -12.673 & -20.039 & -19.335 \\
& FF15--FF16 &
\best{-23.035} & -19.905 & -18.998 & \second{-21.436} & -17.186 & -16.473 &
\best{-23.515} & -20.324 & -21.297 & \second{-23.136} & -15.116 & -15.753 & -17.718 & -19.034 \\
& DF5--DF6 &
\best{-22.197} & -18.389 & -15.099 & \second{-19.605} & -15.012 & -12.258 &
\textemdash & \textemdash & \textemdash & \textemdash & \textemdash & \textemdash & \textemdash & \textemdash \\
& PF10--PF15 &
\best{-12.011} & -6.736 & -5.382 & \second{-8.519} & -7.959 & -7.075 &
\best{-17.546} & -9.144 & -8.888 & -9.899 & \second{-15.783} & -14.192 & -9.591 & -9.844 \\
& WF11--WF25 &
\best{-17.869} & \second{-9.705} & -9.467 & -9.295 & -4.404 & -8.285 &
\textemdash & \textemdash & \textemdash & \textemdash & \textemdash & \textemdash & \textemdash & \textemdash \\
& \textbf{Avg.} &
\best{-19.537} & -14.937 & -12.842 & \second{-16.091} & -12.535 & -12.651 &
\best{-21.394} & -16.760 & -17.097 & \second{-18.901} & -15.466 & -14.206 & -15.783 & -16.071 \\
\bottomrule
\end{tabular}}
\vspace{-1em}
\end{table*}

\subsubsection{Downstream Task Evaluation}

Table~\ref{tab:CP_results} compares the channel prediction performance of the proposed method and representative baselines, where the prediction ratio is set to 25\%. We adopt NMSE (in dB) as the evaluation metric, and lower values indicate more accurate channel reconstruction. Notably, the pre-trained WiFo-MiSAC can be directly applied to all sub-datasets, including OOD ones, without fine-tuning, whereas the task-specific baselines require separate training and evaluation for each sub-dataset. Benefiting from cross-modal masked reconstruction pre-training on the constructed large-scale multimodal sensing-and-communication dataset, WiFo-MiSAC consistently achieves the best performance across all scenarios, including both in-distribution (ID) and out-of-distribution (OOD) splits. Compared with the strongest competing baseline, WiFo-MiSAC improves the averaged NMSE by 2.938/2.200~dB on ID and 3.409/2.700~dB on OOD under unimodal and multimodal inputs, respectively. Moreover, the WiFo-based unimodal foundation model requires additional fine-tuning to handle multimodal inputs; consequently, similar to task-specific multimodal baselines (e.g., Trans-MM), it needs to be trained 134 times, the total number of sub-dataset cases in Table~\ref{tab:CP_results} when both unimodal and multimodal input settings are considered, to cover different modality configurations and system settings. In contrast, WiFo-MiSAC relies on a single unified model to accommodate heterogeneous configurations and sensor combinations, substantially reducing the overhead of model management and switching in practical deployments.

Similarly, Table~\ref{tab:CE_results} reports the channel estimation results with pilot spacing $N_s = 4$, where NMSE (in dB) is also used as the metric. Under the same setting, the pre-trained WiFo-MiSAC can still be directly generalized to all sub-datasets, including OOD ones, without fine-tuning, while the task-specific baselines require separate training and evaluation on each sub-dataset. WiFo-MiSAC again delivers the best overall performance, surpassing the second-best method by 2.615/1.085~dB on ID and 3.446/2.493~dB on OOD for unimodal and multimodal inputs, respectively. These gains further demonstrate its strong reconstruction capability and its flexibility in handling diverse channel reconstruction tasks.

Table~\ref{tab:sensing_beam_results}(a)--(d) summarizes the sensing-oriented and beam-prediction downstream results. Table~\ref{tab:sensing_beam_results}(a) and (b) report distance and AoA estimation on PC9, PC11, and FC15, where WiFo-MiSAC achieves the lowest MAE under all modality configurations. Table~\ref{tab:sensing_beam_results}(c) reports beam prediction on PC11, while Table~\ref{tab:sensing_beam_results}(d) evaluates radar-aided beam prediction on the measured DeepSense~6G datasets (S4--S6). WiFo-MiSAC consistently attains the highest Top-$k$ accuracy, and its advantage over the Scratch baseline highlights the benefit of multimodal pre-training. The results on DeepSense~6G further demonstrate the transferability of the learned representations to real-world measured environments.
\RevisedText{Table~\ref{tab:sensing_beam_results}(e) further evaluates cross-domain generalization through wireless localization on the unseen measured Dichasus dataset~\cite{dichasus2021}. We follow the same dataset configuration and downstream adaptation protocol as WiFo-CF~\cite{liu2025wifo}, and report the published WiFo-CF results for comparison. With only 1{,}000 labeled samples, WiFo-MiSAC establishes new state-of-the-art performance in both indoor and outdoor scenarios, achieving RMSEs of 0.85~m and 2.09~m, respectively. Its consistent improvement over WiFo-CF, despite using only CSI for downstream localization, indicates that multimodal pre-training learns richer and more transferable channel representations under cross-frequency and cross-configuration distribution shifts.}

\begin{table*}[htbp]
\centering
\caption{Evaluation of WiFo-MiSAC across downstream sensing, beam prediction, wireless localization, and RGB modality-extension tasks. The best and second-best results are highlighted in \best{bold} and \second{underlined}, respectively.}
\label{tab:sensing_beam_results}
\setlength{\tabcolsep}{2.6pt}
\renewcommand{\arraystretch}{1.08}
\scriptsize

\begin{minipage}[t]{0.49\textwidth}
\centering
\textbf{(a) Distance estimation (MAE in meters)}
\vspace{0.25em}
\resizebox{\linewidth}{!}{
\begin{tabular}{l ccc ccc ccc}
\toprule
& \multicolumn{3}{c}{Radar-only} & \multicolumn{3}{c}{CSI-only} & \multicolumn{3}{c}{Radar \& CSI} \\
\cmidrule(lr){2-4}\cmidrule(lr){5-7}\cmidrule(lr){8-10}
Model & PC9 & PC11 & FC15 & PC9 & PC11 & FC15 & PC9 & PC11 & FC15 \\
\midrule
WiFo-MiSAC & \best{1.085} & \best{0.454} & \best{1.264} & \best{0.442} & \best{0.527} & \best{0.632} & \best{0.379} & \best{0.377} & \best{0.585} \\
WiFo-based & \second{1.222} & 0.478 & 5.483 & \second{0.515} & \second{0.562} & 0.814 & 0.465 & \second{0.388} & \second{0.956} \\
Scratch & 1.250 & \second{0.466} & \second{4.266} & 0.962 & 1.244 & \second{0.790} & \second{0.418} & 1.481 & 2.212 \\
WiT & 13.47 & 5.002 & 6.805 & 1.830 & 5.128 & 8.232 & 1.840 & 5.092 & 11.88 \\
\bottomrule
\end{tabular}}
\end{minipage}
\hfill
\begin{minipage}[t]{0.49\textwidth}
\centering
\textbf{(b) AoA estimation (MAE in radians)}
\vspace{0.25em}
\resizebox{\linewidth}{!}{
\begin{tabular}{l ccc ccc ccc}
\toprule
& \multicolumn{3}{c}{Radar-only} & \multicolumn{3}{c}{CSI-only} & \multicolumn{3}{c}{Radar \& CSI} \\
\cmidrule(lr){2-4}\cmidrule(lr){5-7}\cmidrule(lr){8-10}
Model & PC9 & PC11 & FC15 & PC9 & PC11 & FC15 & PC9 & PC11 & FC15 \\
\midrule
WiFo-MiSAC & \best{0.479} & \best{0.276} & \best{0.109} & \best{0.081} & \best{0.079} & \best{0.041} & \best{0.053} & \best{0.041} & \best{0.002} \\
WiFo-based & \second{0.493} & 0.299 & \second{0.458} & 0.144 & \second{0.082} & 0.069 & 0.117 & 0.069 & 0.322 \\
Scratch & 0.497 & \second{0.295} & 1.092 & \second{0.118} & 0.483 & \second{0.049} & \second{0.075} & \second{0.049} & 0.089 \\
WiT & 1.249 & 1.405 & 0.725 & 0.956 & 1.414 & 0.050 & 0.867 & 1.404 & \second{0.035} \\
\bottomrule
\end{tabular}}
\end{minipage}

\vspace{0.75em}

\begin{minipage}[t]{0.49\textwidth}
\centering
\textbf{(c) Beam prediction on PC11 (Top-$k$ accuracy)}
\vspace{0.25em}
\resizebox{\linewidth}{!}{
\begin{tabular}{l ccc ccc ccc}
\toprule
& \multicolumn{3}{c}{Radar-only} & \multicolumn{3}{c}{CSI-only} & \multicolumn{3}{c}{Radar \& CSI} \\
\cmidrule(lr){2-4}\cmidrule(lr){5-7}\cmidrule(lr){8-10}
Model & Top-5 & Top-3 & Top-1 & Top-5 & Top-3 & Top-1 & Top-5 & Top-3 & Top-1 \\
\midrule
WiFo-MiSAC & \best{1.000} & \best{0.992} & \best{0.692} & \best{0.998} & \best{0.966} & \best{0.728} & \best{1.000} & \best{1.000} & \best{0.913} \\
WiFo-based & \second{0.978} & \second{0.861} & \second{0.463} & \second{0.992} & \second{0.959} & \second{0.721} & \best{1.000} & \second{0.984} & 0.801 \\
Scratch & 0.892 & 0.729 & 0.365 & 0.990 & 0.933 & 0.627 & 0.896 & 0.681 & 0.500 \\
DNN & 0.859 & 0.611 & 0.312 & 0.913 & 0.790 & 0.504 & \best{1.000} & \second{0.984} & \second{0.869} \\
\bottomrule
\end{tabular}}
\end{minipage}
\hfill
\begin{minipage}[t]{0.49\textwidth}
\centering
\textbf{(d) Radar-aided beam prediction on DeepSense 6G (Top-$k$ accuracy)}
\vspace{0.25em}
\resizebox{\linewidth}{!}{
\begin{tabular}{l ccc ccc ccc}
\toprule
& \multicolumn{3}{c}{S4} & \multicolumn{3}{c}{S5} & \multicolumn{3}{c}{S6} \\
\cmidrule(lr){2-4}\cmidrule(lr){5-7}\cmidrule(lr){8-10}
Model & Top-5 & Top-3 & Top-1 & Top-5 & Top-3 & Top-1 & Top-5 & Top-3 & Top-1 \\
\midrule
WiFo-MiSAC & \best{0.952} & \best{0.932} & \best{0.887} & \best{0.948} & \best{0.930} & \best{0.900} & \best{0.969} & \best{0.929} & \best{0.872} \\
WiFo-based & 0.905 & 0.832 & 0.608 & 0.866 & 0.726 & 0.484 & \second{0.943} & 0.895 & 0.790 \\
Scratch & \second{0.924} & \second{0.907} & \second{0.868} & \second{0.915} & \second{0.903} & \second{0.894} & 0.934 & \second{0.920} & \second{0.864} \\
DNN & 0.593 & 0.446 & 0.245 & 0.544 & 0.394 & 0.208 & 0.688 & 0.498 & 0.169 \\
\bottomrule
\end{tabular}}
\end{minipage}

\vspace{0.75em}

\begin{minipage}[t]{0.49\textwidth}
\centering
\textbf{\RevisedText{(e) Wireless localization on Dichasus (all metrics in meters)}}
\vspace{0.25em}
\setlength{\tabcolsep}{2.2pt}
\renewcommand{\arraystretch}{1.1}
\resizebox{\linewidth}{!}{
\begin{tabular}{l ccc ccc}
\toprule
\multirow{2}{*}{Method}
& \multicolumn{3}{c}{Indoor}
& \multicolumn{3}{c}{Outdoor} \\
\cmidrule(lr){2-4}\cmidrule(lr){5-7}
& L1 Error & L2 Error & P95-L2
& L1 Error & L2 Error & P95-L2 \\
\midrule
WiFo-MiSAC
& \best{1.07} & \best{0.85} & \best{1.86}
& \best{2.44} & \best{2.09} & \best{5.87} \\
WiFo-CF
& \second{1.44} & \second{1.11} & \second{2.39}
& \second{2.85} & \second{2.40} & \second{6.61} \\
Scratch
& 1.78 & 1.37 & 2.77
& 3.64 & 3.05 & 7.31 \\
WiT
& 2.27 & 1.91 & 3.54
& 3.76 & 3.14 & 7.91 \\
\bottomrule
\end{tabular}}
\end{minipage}%
\hfill
\begin{minipage}[t]{0.49\textwidth}
\centering
\textbf{\RevisedText{(f) RGB modality expansion for channel prediction (NMSE)}}
\vspace{0.25em}
\resizebox{\linewidth}{!}{
\begin{tabular}{l ccc ccc}
\toprule
\multirow{2}{*}{Method}
& \multicolumn{3}{c}{FC15}
& \multicolumn{3}{c}{FC16} \\
\cmidrule(lr){2-4}\cmidrule(lr){5-7}
& CSI & CSI\&RGB & Gain
& CSI & CSI\&RGB & Gain \\
\midrule
WiFo-MiSAC
& \best{-16.289} & \best{-17.595} & \best{1.305}
& \best{-18.928} & \best{-19.957} & \best{1.029} \\
WiFo-based
& \second{-11.331} & \second{-11.759} & \second{0.428}
& \second{-12.351} & \second{-12.480} & 0.129 \\
Scratch
& -0.799 & -0.894 & 0.095
& -1.302 & -1.403 & 0.101 \\
Transformer
& -1.238 & -1.475 & 0.237
& -1.599 & -1.778 & \second{0.179} \\
\bottomrule
\end{tabular}}
\end{minipage}

\end{table*}

\subsubsection{Modality Expansion}

In practical deployments, sensor configurations may evolve over time, requiring the model to incorporate new modalities efficiently. We evaluate modality expansion on channel prediction and channel estimation under unseen scenarios. For a \emph{seen} modality, i.e., Radar, WiFo-MiSAC directly reuses the pre-trained Radar tokenizer for zero-shot inference. For \emph{unseen} modalities, i.e., GPS and RGB, WiFo-MiSAC only fine-tunes a lightweight modality tokenizer to map the new measurements into the shared token space, while keeping the multimodal Transformer backbone frozen. Since no modality-specific experts are pre-trained for these new modalities, their tokens are processed through the shared pathway. \RevisedText{This is particularly suitable for the few-shot adaptation setting considered here, because introducing a new modality-specific expert group would add many private parameters that cannot be reliably trained with limited data. By contrast, the shared pathway allows the new modality to directly leverage the environment- and propagation-related knowledge learned during multimodal pre-training.} For comparison, WiFo-based adopts connector-based adaptation~\cite{cheng2025foundation}, while the task-specific Transformer baseline is re-trained for each expanded modality configuration.

\begin{figure*}[htbp]
    \centering
    \includegraphics[width=0.95\linewidth]{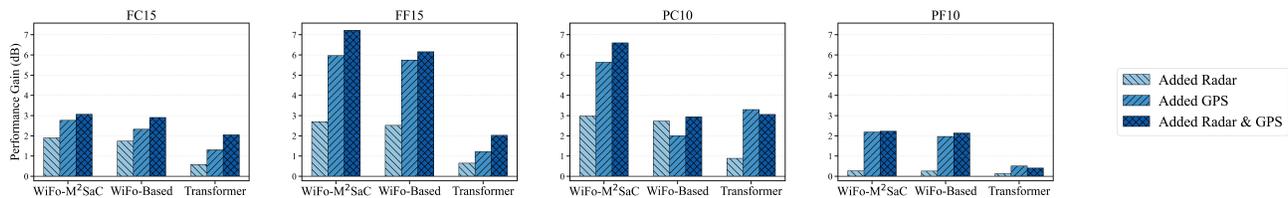}
    \caption{Performance gain of different methods after modality expansion (higher is better).}
    \label{pic:AddModal}
    \vspace{-1em}
    \end{figure*}

Fig.~\ref{pic:AddModal} reports the performance gains over the CSI-only setting after introducing Radar, GPS, or Radar+GPS. WiFo-MiSAC consistently benefits from the added modalities and achieves larger gains than the baselines. In particular, Radar can be exploited in a zero-shot manner, while GPS can be incorporated through lightweight adaptation, demonstrating the flexibility of the proposed expansion strategy. \RevisedText{To further examine whether such expansion ability extends to high-dimensional modalities, we additionally evaluate RGB-assisted channel prediction, as shown in Table~\ref{tab:sensing_beam_results}(f). The results show that adding RGB images brings clear performance gains to WiFo-MiSAC, whereas the improvements of WiFo-based and task-specific baselines are much more limited. This indicates that WiFo-MiSAC can efficiently exploit not only low-dimensional GPS measurements but also high-dimensional visual observations.}These results indicate that the proposed multimodal Transformer, together with the pre-training strategy, learns a well-structured and readily alignable representation space, thereby enabling efficient integration of newly introduced modalities. From a deployment perspective, a single unified model can flexibly support heterogeneous sensor combinations and system configurations, substantially reducing the cost of repeatedly training, maintaining, and switching multiple configuration-specific models in real-world deployments.

\subsubsection{Robustness to Modality Missingness}

\begin{figure*}[htbp]
    \centering
    \includegraphics[width=0.95\linewidth]{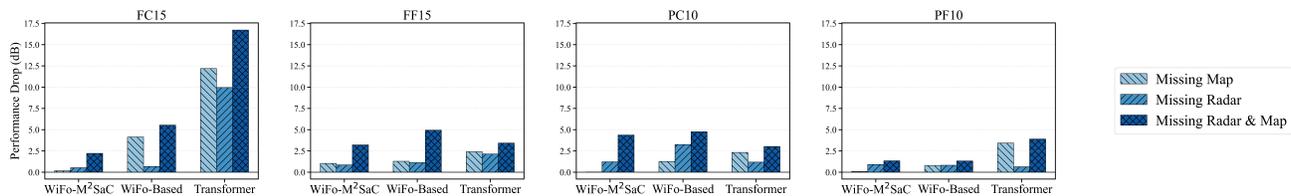}
    \caption{Performance drop of different methods after modality missingness (lower is better).}
    \label{pic:MissModal}
    \vspace{-1em}
    \end{figure*}

We assess robustness to modality missingness by removing one or more modalities during evaluation and reporting the resulting \emph{performance drop} (dB) relative to the full-modality input. Fig.~\ref{pic:MissModal} presents results on four datasets: FC15 and PC10 for channel prediction, and FF15 and PF10 for channel estimation, all evaluated under a 1/4 mask rate. For each dataset, we evaluate three missing-modality settings: \textit{Missing Map}, \textit{Missing Radar}, and \textit{Missing Radar \& Map}. All methods are evaluated without additional fine-tuning: missing modalities are directly removed at inference time, and each model must handle the resulting incomplete inputs using its original parameters.

WiFo-MiSAC consistently exhibits the smallest degradation across all datasets and missing-modality settings, with an average performance drop of only 1.33\,dB, whereas the task-specific model degrades by 5.11\,dB on average. This robustness can be attributed to both the model design and the pre-training strategy. Architecturally, WiFo-MiSAC explicitly decouples modality-specific representations from shared representations, reducing cross-modality dependency and thereby limiting the negative impact of missing or corrupted modalities on the remaining inputs. In addition, the pre-training dataset contains a large fraction of incomplete modality observations, and the modality dropout applied during pre-training further regularizes the model to operate reliably under missing-modality conditions. Overall, these results confirm that WiFo-MiSAC is substantially more robust to incomplete sensor observations, which is essential for real-world deployments where sensor failures, occlusions, and intermittent availability are common.

\subsubsection{Ablation Experiments}

\begin{figure}[htbp]
    \centering
    \includegraphics[width=1\linewidth]{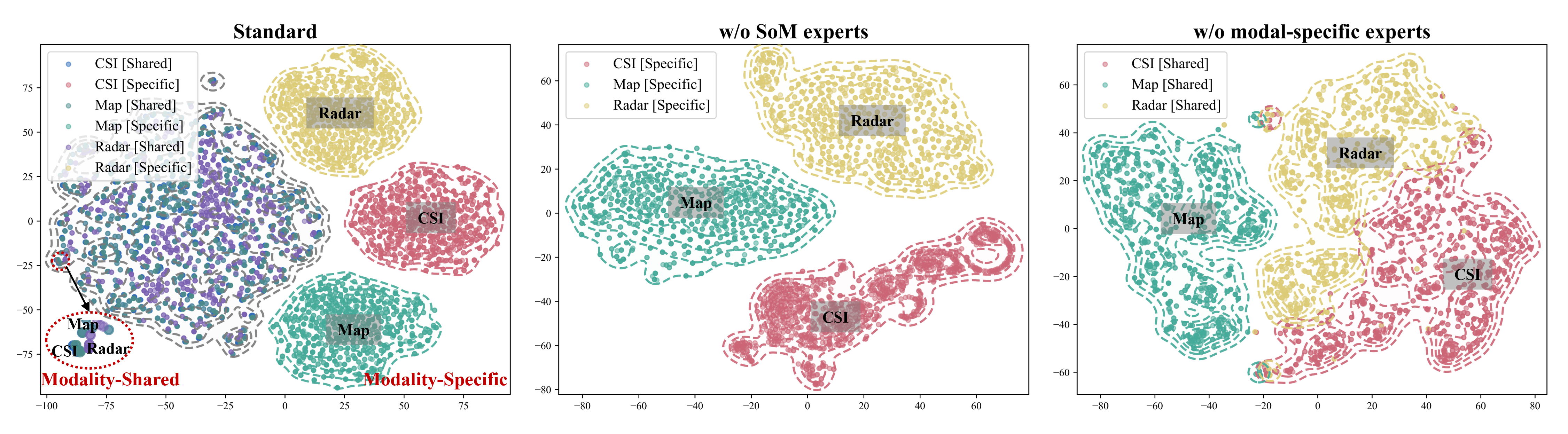}
    \caption{The t-SNE visualization of CSI, Radar, and Map features extracted by three WiFo-MiSAC variants on PF5.}
    \label{fig:tsne}
    \end{figure}

To validate the effectiveness of the proposed scheme, we conduct ablation studies on dataset construction, model architecture, and pre-training objectives. Specifically, we remove the map modality from the pre-training dataset (\textit{w/o map data}) and discard unpaired samples (\textit{w/o unpaired data}); we ablate architectural components by removing the shared SoM expert, the modality-specific expert, and the scene-adaptive routing (\textit{w/o SoM expert}, \textit{w/o specific expert}, and \textit{w/o scene-adaptive routing}). \RevisedText{For a fair comparison, in the \textit{w/o SoM expert} and \textit{w/o specific expert} variants, we increase the number of the remaining experts to keep the total model parameter count comparable to that of the standard configuration.} \textcolor{black}{We further investigate the role of pre-training objectives by disabling the contrastive loss and three masked reconstruction strategies, i.e., random masking, frequency-domain masking, and comb masking (\textit{w/o} $\mathcal{L}_{\mathrm{cl}}$, \textit{w/o} $\mathcal{L}^{random}_{\mathrm{mask}}$, \textit{w/o} $\mathcal{L}^{frequency}_{\mathrm{mask}}$, and \textit{w/o} $\mathcal{L}^{comb}_{\mathrm{mask}}$)}. Table~\ref{tab:ablation} reports the results, where CP/BP/DE/AE are evaluated on PC11 and CE is evaluated on PF9. All ablated variants consistently underperform the standard configuration, confirming the effectiveness of each component in WiFo-MiSAC. Notably, the degradation caused by removing different masking strategies demonstrates that masked reconstruction is a key factor in learning robust and transferable multimodal representations.

To further illustrate the benefit of explicitly decoupling modality-specific and shared experts, we visualize the extracted multimodal representations using t-SNE for three variants: the standard model, \textit{w/o SoM expert}, and \textit{w/o modality-specific expert}, as shown in Fig.~\ref{fig:tsne}. In the standard setting, modality-specific features form three clearly separated clusters, while the shared features (blue-toned points) exhibit paired aggregation across modalities, indicating that the model captures both modality-unique characteristics and modality-invariant shared factors with strong interpretability. Without the SoM expert, representations are dominated by modality-specific clusters, and the shared structure across modalities becomes weak, suggesting limited ability to learn cross-modal common representations and thus reduced cross-modal complementarity. In contrast, removing modality-specific experts forces all modalities to share a single representation space, leading to entangled and disorganized distributions; this indicates difficulty in simultaneously modeling commonality and specificity across modalities, which explains the performance degradation observed in Table~\ref{tab:ablation}.

\begin{table}[htbp]
\centering
\caption{\RevisedText{Ablation results for WiFo-MiSAC downstream tasks with multimodal inputs. CP and CE tasks utilize a 1/4 mask rate.}}
\label{tab:ablation}
\setlength{\tabcolsep}{2pt}
\renewcommand{\arraystretch}{1.15}
\resizebox{\columnwidth}{!}{
\begin{tabular}{lccccc}
\toprule
\textbf{Variant} &
\makecell[c]{\textbf{CP} \\ (NMSE$\downarrow$)} &
\makecell[c]{\textbf{CE} \\ (NMSE$\downarrow$)} &
\makecell[c]{\textbf{BP} \\ (Top-1$\uparrow$)} &
\makecell[c]{\textbf{DE} \\ (MAE$\downarrow$)} &
\makecell[c]{\textbf{AE} \\ (MAE$\downarrow$)} \\
\midrule
Standard & \textbf{-7.279} & \textbf{-18.640} & \textbf{0.913} & \textbf{0.377} & \textbf{0.041} \\
\midrule
w/o map data & -7.089 & -16.981 & 0.891 & 0.461 & 0.056 \\
w/o unpaired data & -5.287 & -16.531 & 0.884 & 0.440 & 0.053 \\
\midrule
w/o $\mathcal{L}_{\mathrm{cl}}$ & -6.915 & -16.529 & 0.862 & 0.574 & 0.055 \\
w/o $\mathcal{L}^{random}_{\mathrm{mask}}$ & -2.846 & -17.592 & 0.882 & 0.516 & 0.061 \\
w/o $\mathcal{L}^{frequency}_{\mathrm{mask}}$ & -1.031 & -18.581 & 0.894 & 0.498 & 0.059 \\
w/o $\mathcal{L}^{comb}_{\mathrm{mask}}$ & -5.273 & -11.782 & 0.893 & 0.471 & 0.057 \\
\midrule
\RevisedText{w/o SoM expert} & \RevisedText{-6.903} & \RevisedText{-18.153} & \RevisedText{0.889} & \RevisedText{0.495} & \RevisedText{0.048} \\
\RevisedText{w/o specific expert} & \RevisedText{-6.551} & \RevisedText{-16.759} & \RevisedText{0.897} & \RevisedText{0.506} & \RevisedText{0.049} \\

\makecell[c]{w/o scene-adaptive routing} & -6.953 & -17.649 & 0.896 & 0.541 & 0.054 \\
\bottomrule
\end{tabular}}
\vspace{-1em}
\end{table}

\begin{figure}[htbp]
    \centering
    \vspace{-1em}
    \includegraphics[width=\linewidth]{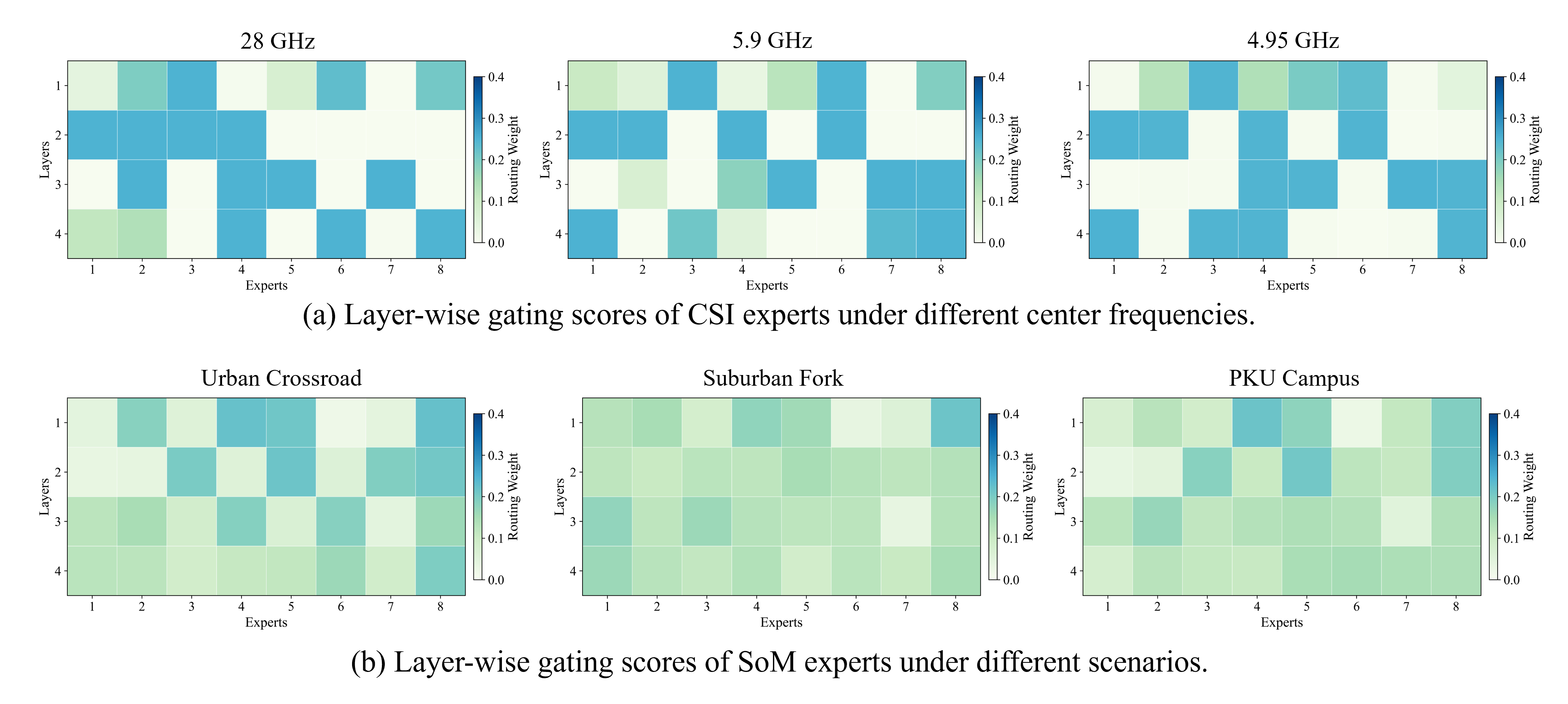}
    \caption{Heatmaps of routing weights for SoM experts and task-specific CSI experts across different scenarios and center frequencies.}
    \label{fig:heatmap}
    \vspace{-1em}
    \end{figure}

\subsubsection{Activation Visualization}
\textcolor{black}{Figure~\ref{fig:heatmap} provides an intuitive view of the routing behavior of the two expert groups. In Fig.~\ref{fig:heatmap}(a), the CSI experts exhibit clearly different activation patterns across input channels with different center frequencies, indicating that the MoE router learns frequency-selective preferences. This behavior suggests that different experts specialize in modeling distinct propagation characteristics under different spectral conditions. By dynamically assigning frequency-dependent inputs to different experts, the model can better accommodate the large distribution gaps across frequency bands, which substantially improves its generalization ability.}

\textcolor{black}{In Fig.~\ref{fig:heatmap}(b), the SoM experts also show scenario-dependent routing patterns under common environments such as \emph{Urban Crossroad}, \emph{Suburban Fork}, and \emph{PKU Campus}. The activated expert combinations vary noticeably with the scene context, demonstrating that the introduced environmental context information effectively guides the router to perform adaptive expert selection. This observation verifies that the proposed contextual design enables the synesthesia experts to capture scene-specific priors and adjust their representations accordingly, thereby enhancing the model's adaptability to diverse deployment scenarios.}

\begin{figure*}[htbp]
    \centering
    \vspace{-1em}
    \includegraphics[width=0.95\linewidth]{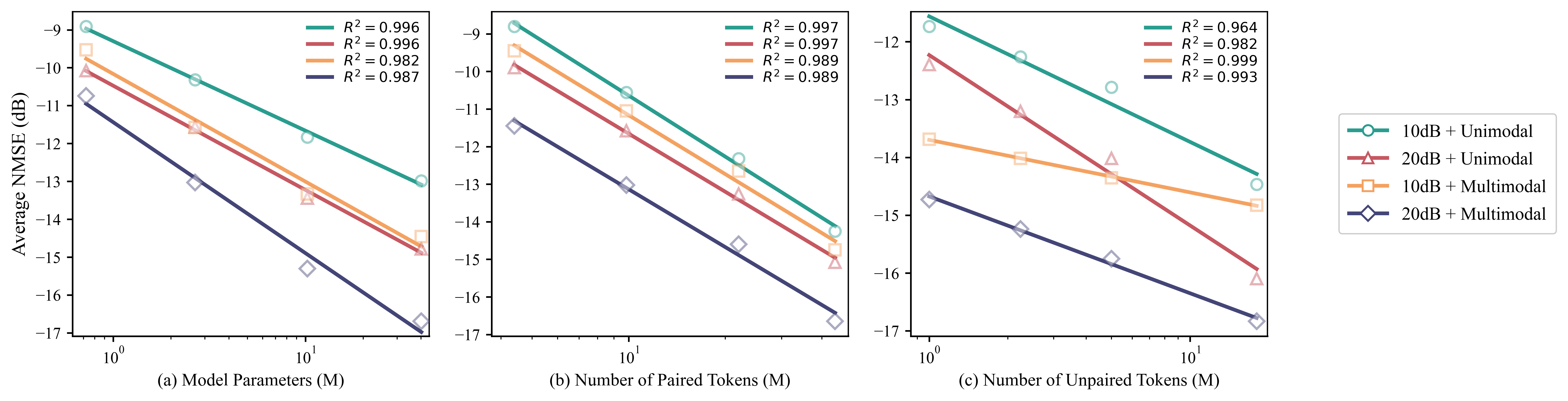}
\caption{\RevisedText{Scaling laws of WiFo-MiSAC for the channel estimation task with a masking ratio of 1/4}}
    \label{fig:scale_law}
    \vspace{-1em}
    \end{figure*} 

\subsubsection{Scaling Law Analysis}
Fig. ~\ref{fig:scale_law} reveals clear scaling trends from both the model-capacity and data-scale perspectives. \RevisedText{Considering the practical deployment constraints of wireless physical-layer systems, we focus on empirical scaling behavior within the sub-100M-parameter regime, instead of extrapolating the observed trends to billion-parameter wireless models \cite{liu2024wifo, yang2025wirelessgpt, alikhani2024large}.} In Fig.~\ref{fig:scale_law}(a), the average NMSE consistently decreases as the model size grows for channel estimation, demonstrating stable model scaling behavior. Moreover, the multimodal curves generally exhibit steeper improvement trends than the unimodal ones, especially in high-SNR settings, suggesting that larger models are better able to exploit cross-modal correlations and translate increased capacity into larger performance gains.

\RevisedText{In Figs.~\ref{fig:scale_law}(b) and (c), increasing the amount of both unpaired and paired tokens yields consistent performance gains. Specifically, the unpaired tokens in Fig.~\ref{fig:scale_law}(c) refer to CSI tokens from the \textit{Dense Building} and \textit{Urban Wide Lane} scenes, where paired observations from other modalities are unavailable, while the paired tokens in Fig.~\ref{fig:scale_law}(b) correspond to time-synchronized CSI--radar--map samples. These results indicate that WiFo-MiSAC can effectively benefit from different data alignment regimes: unpaired CSI data strengthens modality-specific channel representation learning, whereas paired multimodal data further enhances cross-modal alignment and modality-shared environmental representation learning. This is because our architecture explicitly models modality-specific representations from modality-shared representations. When handling missing modalities, the training process mainly emphasizes modality-specific representation learning, while removing the contrastive term to avoid interfering with the shared experts that have acquired cross-modal common representations from paired data. As a result, WiFo-MiSAC preserves favorable scaling behavior under both unpaired and paired data expansion, demonstrating robust and scalable multimodal representation learning.}

\begin{table}[htbp]
\centering
\scriptsize
\caption{Activated / total parameters (M) of WiFo-MiSAC and baselines under different modality inputs.}
\label{tab:params}
\setlength{\tabcolsep}{6pt}
\renewcommand{\arraystretch}{1.15}
\resizebox{\columnwidth}{!}{
\begin{tabular}{lccc}
\toprule
\textbf{Modality} & \textbf{WiFo-MiSAC} & \textbf{WiFo-based} & \textbf{Task-specific}\\
\midrule
CSI   & 17.23 / 27.72 & 17.23 / 27.72 & 7.97 / 7.97 \\
Radar & 13.05 / 21.45 & 13.05 / 21.45 & 8.02 / 8.02
\\
Map   & 13.05 / 21.45 & 13.05 / 21.45 & 8.02 / 8.02 \\
\midrule
All modalities & \textbf{26.35 / 45.24} & 43.33 / 70.62 & 24.01 / 24.01
\\
\bottomrule
\end{tabular}}
\vspace{-1em}
\end{table}

\begin{table}[htbp]
\centering
\caption{FLOPs (G) of WiFo-MiSAC and baselines on five downstream tasks.}
\label{tab:flops}
\setlength{\tabcolsep}{3pt}
\renewcommand{\arraystretch}{1.15}
\resizebox{\columnwidth}{!}{
\begin{tabular}{lccc}
\toprule
\textbf{Downstream task} & \textbf{WiFo-MiSAC} & \textbf{WiFo-based} & \textbf{Task-specific} \\
\midrule
Channel Prediction   & 16.473 & 18.625 & 8.946 \\
Channel Estimation   & 16.473 & 18.625 & 5.607 \\
Beam Prediction      & 14.196 & 16.662 & 4.938 \\
Distance Estimation  & 14.196 & 16.662 & 9.760 \\
AoA Estimation       & 14.196 & 16.662 & 9.760 \\
\midrule
All tasks & \textbf{18.785} & 30.487 & 39.012 \\
\bottomrule
\end{tabular}}
\vspace{-1em}
\end{table}
\subsubsection{Complexity Analysis}
\textcolor{black}{We examine whether WiFo-MiSAC also offers practical efficiency advantages {while retaining its performance gains}.
Therefore, we compare different methods from two complementary perspectives, namely parameter cost and computational cost, as reported in Tables~\ref{tab:params} and~\ref{tab:flops}.}
To ensure a fair architectural match, the \emph{task-specific} baselines in Tables~\ref{tab:params} and~\ref{tab:flops} are instantiated with Transformer-based models whenever applicable to match the Transformer-style backbone of WiFo-MiSAC (e.g., WiT-style backbones for distance/AoA estimation and Transformer predictors for channel prediction/estimation), while beam prediction follows the commonly used BP-DNN baseline.
\textcolor{black}{Table~\ref{tab:params} shows that in addition to achieving superior overall performance, WiFo-MiSAC also reduces the overall parameter cost by sharing self-attention layers and SoM experts across modalities, instead of maintaining separate modality-specific backbones.} Compared with WiFo-based, WiFo-MiSAC reduces the total parameters by {39.2\%} and the activated parameters by {35.9\%}. 

Table~\ref{tab:flops} further reports the FLOPs required by different approaches on multiple downstream tasks. All FLOPs are computed by running inference on the {CC1} dataset with batch size $=8$, \RevisedText{where the CSI dimension is $32\times32$ and the mask rate for channel prediction and estimation is set to $1/4$.} In total, WiFo-MiSAC requires 18.785\,G FLOPs, which is {38.4\%} lower than WiFo-based and {51.8\%} lower than deploying separate task-specific models. 
\textcolor{black}{These results indicate that the performance advantage of WiFo-MiSAC is not obtained at the expense of higher complexity; instead, by enabling modality-level parameter sharing and cross-task representation reuse, it simultaneously delivers stronger effectiveness and better deployment efficiency under heterogeneous sensor configurations.}


\section{Conclusions}

This paper introduced WiFo-MiSAC, a task-agnostic foundation model that unified heterogeneous wireless sensing and communication within a single framework. The SS-DMoE architecture was employed to decouple modality-shared and modality-specific representations, mitigating interference while enabling deep cross-modal interactions. Feature transferability was further enhanced through a hybrid pre-training strategy combining masked reconstruction with contrastive learning. Extensive experiments demonstrated superior performance across diverse downstream tasks, alongside robust handling of missing modalities and efficient expansion to new sensor configurations. Ablation studies and complexity analyses confirmed the efficiency of the unified design, establishing WiFo-MiSAC as a practical backbone for future intelligent wireless systems.

\bibliographystyle{IEEEtran}

\bibliography{WiFo-MiSAC}

\end{document}